\documentclass{article}

\PassOptionsToPackage{backref=page}{hyperref}

\usepackage{microtype}
\usepackage{graphicx}
\usepackage{makecell}
\usepackage{subfigure}
\usepackage{booktabs} 

\usepackage{hyperref}
\usepackage{xspace}
\usepackage{upgreek}
\usepackage{enumitem}

\usepackage[accepted]{icml2025}

\usepackage{amsmath}
\usepackage{amssymb}
\usepackage{capt-of}

\usepackage{mathtools}
\usepackage{amsthm}
\usepackage{thm-restate}
\usepackage{multirow}

\DeclareMathOperator*{\argmax}{arg\,max}
\DeclareMathOperator*{\argmin}{arg\,min}
\newcommand{\bignorm}[1]{\left\lVert#1\right\rVert}
\newcommand{\norm}[1]{\lVert#1\rVert}

\usepackage[capitalize]{cleveref}
\crefname{equation}{}{}
\Crefname{equation}{Eq.}{Eqs.}
\Crefname{figure}{Fig.}{Figs.}
\Crefname{figures}{Figs.}{Figs.}
\Crefname{section}{Sec.}{Secs.}
\Crefname{sections}{Secs.}{Secs.}
\Crefname{table}{Tab.}{Tabs.}
\Crefname{tables}{Tabs.}{Tabs.}
\Crefname{appendix}{App.}{Apps.}

\theoremstyle{plain}
\newtheorem{theorem}{Theorem}[section]
\newtheorem{proposition}[theorem]{Proposition}

\theoremstyle{definition}

\theoremstyle{remark}

\usepackage[textsize=tiny]{todonotes}

\newcommand{\fullname}{Denoising Diffusion Codebook Models\xspace}

\newcommand{\dynamictitle}{Compressed Image Generation with \fullname}

\usepackage{amsfonts,bm}

\def\eqref#1{equation~\ref{#1}}

\def\1{\bm{1}}

\def\rvs{{\mathbf{s}}}

\def\rvw{{\mathbf{w}}}
\def\rvx{{\mathbf{x}}}
\def\rvy{{\mathbf{y}}}
\def\rvz{{\mathbf{z}}}

\def\vzero{{\bm{0}}}

\def\vmu{{\bm{\mu}}}

\def\vr{{\bm{r}}}
\def\vs{{\bm{s}}}

\def\vy{{\bm{y}}}
\def\vz{{\bm{z}}}

\def\mA{{\bm{A}}}

\def\mI{{\bm{I}}}

\def\gC{{\mathcal{C}}}

\def\gL{{\mathcal{L}}}

\def\gN{{\mathcal{N}}}

\def\delt{\mathrm{d}}

\definecolor{green300}{HTML}{60d360}
\definecolor{green500}{HTML}{32b432}
\definecolor{green1000}{HTML}{2ca02c}
\definecolor{greenpursuit}{HTML}{047d50}
\definecolor{bluepsc}{HTML}{1f77b4}
\definecolor{lightbluepsc}{HTML}{aec7e8}

\icmltitlerunning{\dynamictitle}

\begin{document}

\twocolumn[
\icmltitle{
\dynamictitle
}



\icmlsetsymbol{equal}{*}

\begin{icmlauthorlist}
\icmlauthor{Guy Ohayon}{equal,techcs}
\icmlauthor{Hila Manor}{equal,techee}
\icmlauthor{Tomer Michaeli}{techee}
\icmlauthor{Michael Elad}{techcs}
\end{icmlauthorlist}

\icmlaffiliation{techcs}{Faculty of Computer Science, Technion -- Israel Institute of Technology, Haifa, Israel}
\icmlaffiliation{techee}{Faculty of Electrical and Computer Engineering, Technion -- Israel Institute of Technology, Haifa, Israel}

\icmlcorrespondingauthor{Guy Ohayon}{guyoep@gmail.com}
\icmlcorrespondingauthor{Hila Manor}{hila.manor@campus.technion.ac.il}

\icmlkeywords{Machine Learning, ICML}

\vskip 0.3in

]



\printAffiliationsAndNotice{\icmlEqualContribution} 

\begin{abstract}
We present a novel generative approach based on Denoising Diffusion Models (DDMs), which produces high-quality image samples \emph{along} with their losslessly compressed bit-stream representations.
This is obtained by replacing the standard Gaussian noise sampling in the reverse diffusion with a selection of noise samples from pre-defined codebooks of fixed iid Gaussian vectors.
Surprisingly, we find that our method, termed \emph{Denoising Diffusion Codebook Model} (DDCM), retains sample quality and diversity of standard DDMs, even for extremely small codebooks.
We leverage DDCM and pick the noises from the codebooks that best match a given image, converting our generative model into a highly effective lossy image codec achieving state-of-the-art perceptual image compression results.
More generally, by setting other noise selections rules, we extend our compression method to any conditional image generation task (e.g., image restoration), where the generated images are produced jointly with their condensed bit-stream representations.
Our work is accompanied by a mathematical interpretation of the proposed compressed conditional generation schemes, establishing a connection with score-based approximations of posterior samplers for the tasks considered.
Code and demo are available on our project's \href{https://ddcm-2025.github.io/}{website}.
\end{abstract}

\begin{figure*}[t]
    \centering
    \includegraphics[width=1\textwidth]{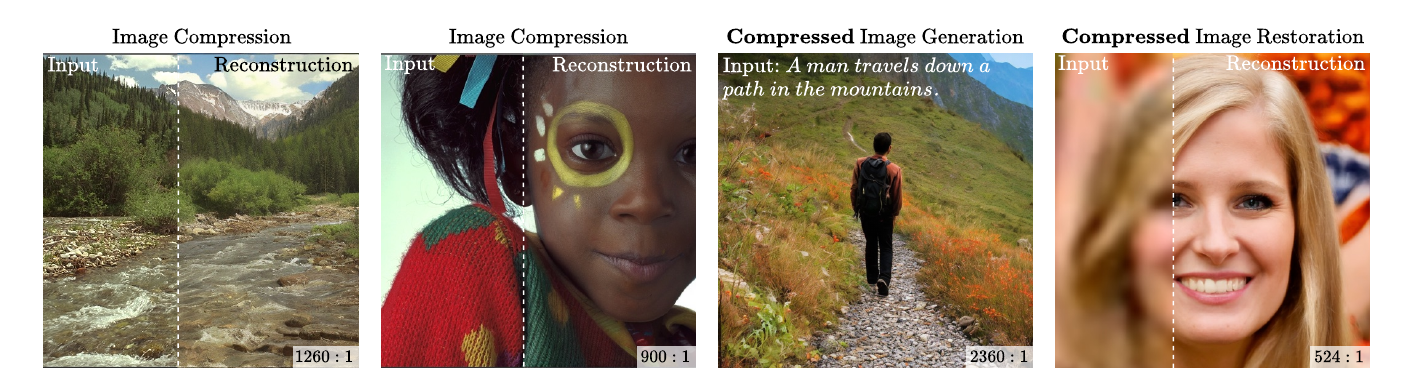}
    \caption{Our proposed scheme (DDCM) produces visually appealing image samples with high compression ratios (bottom-right corners).}
    \label{fig:teaser}
\end{figure*}

\section{Introduction}\label{sec:intro}

Denoising Diffusion Models (DDMs)~\citep{sohl2015deep, ho2020denoising} have emerged as an effective tool for generating samples from complex signal distributions (e.g., natural images).
Hence, DDMs are commonly leveraged to facilitate a variety of downstream tasks, such as text-to-image synthesis~\cite{ramesh2021zero,rombach2022high,saharia2022photorealistic},
editing~\citep{meng2022sdedit,huberman2024edit}, compression~\cite{theis2022lossy, elata2024zero, korber2024perco}, and restoration~\citep{kawar2022denoising, chung2023diffusion}. 
Common to many of these applications is the reliance on iterative sampling from a continuous Gaussian distribution, yielding an unbounded representation space.

This work embarks on the hypothesis that such an infinite representation space is highly redundant.
For example, consider any stochastic diffusion generative process with $T=1000$ sampling steps (e.g., DDPM~\citep{ho2020denoising}).
Suppose that at each timestep, the generative process is restricted to choosing between only two fixed noise realizations.
Sampling could then lead to $2^{1000}$ different outputs, an incredibly large number exceeding the estimated amount of atoms in the universe.
Thus, in principle, such a process could cover the distribution of natural images densely.

\begin{figure*}[t]
    \centering
    \includegraphics[width=\linewidth]{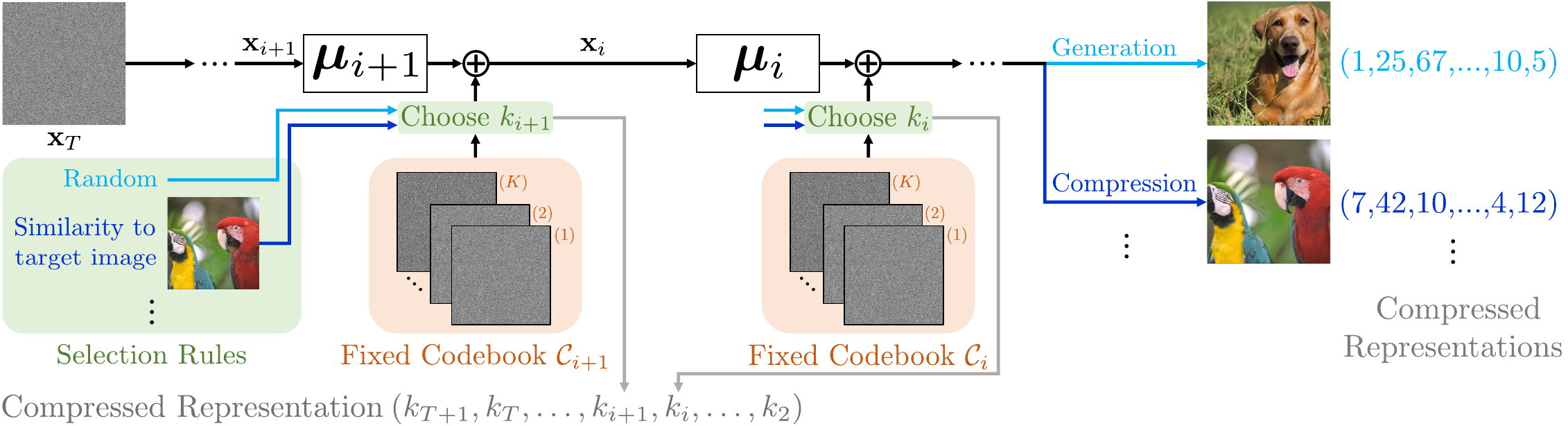}
    \caption{\textbf{Method overview.} 
    DDCM replaces the standard Gaussian noises in DDPM sampling with a selection of noise samples from pre-defined codebooks of fixed iid Gaussian vectors.
    This retains the high-quality generative properties of standard DDMs, while producing the results along with their compressed representations.
    By choosing the discrete noises according to different selection rules, DDCM can perform a variety of conditional image generation tasks. 
    Our highly condensed bit-stream representation is especially effective for image compression, leading to state-of-the-art results.
    }
    \label{fig:overview}
\end{figure*}

We harness this intuition and propose \emph{Denoising Diffusion Codebook Models} (DDCM), 
a novel DDM generation scheme for continuous signals, leveraging a discrete and finite representation space. 
In particular, we first construct a chain of \emph{codebooks}, where each is a sequence of pre-sampled Gaussian noise vectors.
These codebooks are constructed once and remain fixed for the entire lifetime of the model.
Then, during the generative process, we simply randomly pick the noises from the codebooks instead of drawing them from a Gaussian distribution, as shown in \Cref{fig:overview}.
Since we alter only the sampling process, DDCM can be applied using any pre-trained DDM.
Interestingly, we find that our proposed discrete and finite representation space is indeed expressive enough to retain the generative capabilities of standard DDMs, even when using incredibly small codebooks.
Since our generative process is entirely governed by the noise \emph{indices} picked during the generation, an important consequence is that every generated image can be perfectly reconstructed by repeating the process with its corresponding indices.

We leverage this property to solve a variety of tasks, using gradient-free noise selection rules to guide the DDCM generation process.
In particular, by choosing the discrete noises to best match a given image, we achieve state-of-the-art perceptual compression results.
Moreover, using DDCM with different noise selection rules yields a versatile framework for other \emph{compressed} conditional generation tasks, such as compressed image restoration (see examples in \Cref{fig:teaser}).
Finally, we provide a mathematical interpretation of the proposed schemes based on score-based generative modeling with SDEs~\citep{song2020score}, showing a connection between our generalized selection rules and approximate posterior sampling for compressed conditional generation.

\section{Related Work}\label{sec:related}

\paragraph{Compression.}
Image compression has seen significant progress in recent years, with the penetration of neural networks to this domain. Neural methods range from constructing specialized architectures~\citep{end-to-end-compression,zhu2022unified, jiang2023mlic,jiang2023mlicpp} to relying on different generative models such as Generative Adversarial Networks (GANs)~\citep{mentzer2020high,muckley2023improving, iwai2024controlling} and Variational Autoencoders (VAEs)~\citep{theis2017lossy}.
Recent compression methods leverage DDMs and offer high perceptual quality results, by training models from scratch~\citep{NEURIPS2023_ccf6d8b4, ghouse2023residual}, fine-tuning existing models~\cite{careil2023towards,korber2024perco}, or using pre-trained DDMs in a zero-shot manner (without further training)~\citep{theis2022lossy,elata2024zero}. 
Current solutions in the latter category are highly computationally demanding, either due to their communication schemes (e.g., reverse channel coding~\citep{pmlr-v162-theis22a,theis2022lossy}) or their need to perform thousands of denoising operations~\citep{elata2024zero}. Our work falls into this last category, offering a novel and highly effective compression scheme with a fast bit-stream communication method and computational demands that match standard use of DDMs.

\paragraph{Discrete Generative Modeling for Continuous Data.}
Recent works have explored discrete generative modeling of continuous data distributions.
These employ various discrete representations, such as vector quantized latent tokens~\citep{wu2024rdpm} or hierarchical modeling schemes that gradually refine each generated sample~\citep{yang2024discretedistributionnetworks}. DDCM offers a new such discrete generative framework, building on the exceptional achievements of DDMs.

\paragraph{Conditional Image Generation with Pre-Trained DDMs.}
Pre-trained DDMs are commonly utilized for solving conditional image generation tasks, such as image restoration~\citep{kawar2022denoising,lugmayr2022repaint,wang2023zeroshot, chung2023diffusion,2023diffbir, song2023pseudoinverseguided,cohen2024posterior, difface,raphaeli2025silosolvinginverseproblems,man2025proxiesdistortionconsistencyapplications}
and editing~\citep{meng2022sdedit, huberman2024edit, cohen2024slicedit, manor2024zeroshot}. 
In this work we address these tasks from the lens of DDCM, where the conditional samples are generated along with their compressed bit-streams.

\paragraph{Compressed Image Generation.}The task of compressed image generation (generating images directly in their compressed form) has been previously explored. \citet{kang2019jointimagegenerationcompression} trained an unconditional GAN~\citep{NIPS2014_5ca3e9b1} to synthesize JPEG representations. \citet{bulla2023} trained a text-conditional GAN in the JPEG domain.
In this work, we use DDMs instead of GANs and introduce a novel compressed image representation space different than that of JPEG. Our approach is compatible with any pre-trained DDM without requiring additional training.

\paragraph{Compressed Conditional Image Generation.}
Our work also addresses the problem of compressed \emph{conditional} generation (see~\cref{sec:compressed_conditional_generation}), in which the compressed output is generated in accordance with a given input condition, such as a text prompt, a noisy image, or other forms of guidance.
In this context, \citet{liu2021lossy} proposed a method based on optimal transport, subject to an informational bottleneck constraint.
Specifically, their approach involves computing an optimal transport map from the distribution of input conditions to the distribution of target signals, while constraining the entropy of an intermediate latent random variable (i.e., the code).
This method requires both the inputs and outputs to reside within the same metric space.
In contrast, our approach naturally accommodates more complex types of input-output pairs, such as those involving different modalities (e.g., text and image).
\section{Background}
Diffusion models~\citep{sohl2015deep,ho2020denoising,song2020score} generate samples from a data distribution $p_{0}$ by \emph{reversing} a diffusion process that gradually adds random noise to samples from the data.
Specifically, the diffusion process starts with $\rvx_{0}\sim p_{0}$ and produces the chain $\rvx_{0},\rvx_{1},~\hdots~,\rvx_{T}$
via
\begin{align}
&\rvx_{i}=\sqrt{\alpha_{i}}\rvx_{i-1}+\sqrt{1-\alpha_{i}}\rvz_{i},\quad\rvz_{i}\sim\mathcal{N}(\vzero,\mI),\label{eq:forward_gaussian_diffusion}
\end{align}
where $\alpha_{1},\hdots,\alpha_{T}>0$ are some time-dependent constants.
The above is a time-discretization of a Variance Preserving (VP) SDE~\citep{song2020score}.
Then, samples from the data distribution $p_{0}$ are generated by solving the corresponding reverse-time VP SDE~\citep{ANDERSON1982313,song2020score}, i.e., by gradually \emph{denoising} samples, starting from $\rvx_{T}\sim\mathcal{N}(\vzero,\mI)$.
In this paper we adopt Denoising Diffusion Probabilistic Models (DDPMs)~\citep{ho2020denoising}, which propose the generative process ($i=T,\ldots,1$)
\begin{align}
    &\rvx_{i-1}=\vmu_{i}(\rvx_{i})+\sigma_{i}\rvz_{i},~\mbox{where}\label{eq:ddpm}\\
    &\vmu_{i}(\rvx_{i})=\frac{1}{\sqrt{\alpha_{i}}}(\rvx_{i}+(1-\alpha_{i})\vs_{i}(\rvx_{i})),
\end{align}
$\rvz_{i}\sim\mathcal{N}(\vzero,\mI)$, $\sigma_{i}=\sqrt{1-\alpha_{i}}$, and $\vs_{i}(\rvx_{i})$ denotes the \emph{score} $\nabla_{\rvx_{i}}\log{p_{i}(\rvx_{i})}$ of the probability density function $p_{i}(\rvx_{i})$.
Such a score function $\vs_{i}(\rvx_{i})$ is typically learned via \emph{denoising score matching}~\citep{dsm,NEURIPS2019_3001ef25,song2020score,ho2020denoising}, where a model $\hat{\rvx}_{0|i}$ is trained to predict $\rvx_{0}$ from $\rvx_{i}$ (i.e., a denoiser), and using the well-known equation~\citep{robbins1956empirical,miyasawa1961empirical, stein1981estimation}
\begin{align}\label{eq:x0eq}
\vs_{i}(\rvx_{i})=\frac{\sqrt{\bar{\alpha}_{i}}\hat{\rvx}_{0|i}-\rvx_{i}}{1-\bar{\alpha}_{i}},
\end{align}
where $\bar{\alpha}_{i}\coloneqq\prod_{s=1}^{i}\alpha_{s}$.
This generative process is applicable to both pixel space~\citep{dhariwal2021diffusion} and latent space~\citep{rombach2022high} diffusion models, by employing a VAE-based encoder-decoder.

\section{Denoising Diffusion Codebook Models}\label{sec:method}
\paragraph{Method.}Equation~(\ref{eq:ddpm}) depicts the standard DDPM sampling approach, where the added noise is sampled from a continuous Gaussian distribution.
DDCM instead uses a discrete noise space, by limiting each sampling step to choose from $K$ constant noise realizations, fixed separately for each step.
Formally, for each $i=2,\ldots,T+1$ we define a codebook of $K$ entries
\begin{align}
   \gC_i = \left[\vz_i^{(1)},\vz_i^{(2)}, \ldots, \vz_i^{(K)}\right],
\end{align}
where each $\gC_{i}(k)\coloneqq\vz_{i}^{(k)}$ is sampled independently from $\mathcal{N}(\vzero,\mI)$ and remains fixed throughout the model's lifetime.
Then, we modify the DDPM sampling process (\ref{eq:ddpm}), replacing the noise $\rvz_{i}$ by a randomly selected codebook entry,
\vspace{-0.19em}
\begin{align}\label{eq:DDCM_sampling}
    \rvx_{i-1} = \vmu_i(\rvx_i) + \sigma_i \gC_i(k_i),
\end{align}
where the process is initialized with $\rvx_{T}=\gC_{T+1}(k_{T+1})$, $k_i\sim\text{Unif}(\{1, \ldots, K\})$, and sampling step $i=1$ does not involve noise addition. This random selection procedure is depicted in the generation branch in \Cref{fig:overview}.
Importantly, running the generative process~(\ref{eq:DDCM_sampling}) with a given sequence of noise vectors $\{\gC_{i}(k_i)\}_{i=2}^{T+1}$ always produces the same output image.
Thus, as depicted in the bottom part of \Cref{fig:overview}, the sequence of indices $k_{T+1},\hdots,k_2$ can be considered a losslessly compressed bit-stream representation of each generated image. 

\begin{figure}[t]
    \centering
    \includegraphics[width=\linewidth]{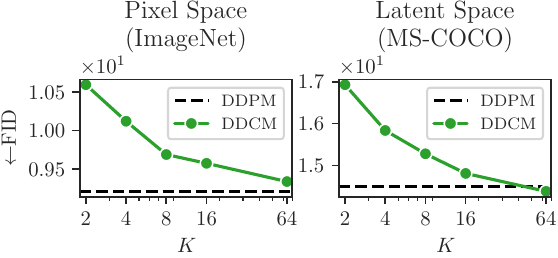}\vspace{-0.3cm}
    \caption{\textbf{Comparing DDPM with DDCM for different codebook sizes $K$.}
    Interestingly, DDCM with $K=64$ achieves similar FID to DDPM, suggesting that the continuous representation space of DDPM (DDCM with $K=\infty$) is highly redundant.
    We use a class-conditional ImageNet model ($256\times256$) for pixel space, and the text-conditional SD 2.1 model ($768\times768$) for latent space, with prompts from MS-COCO.
    The $K$ axis is in log-scale.
}    
    \label{fig:generation_fid_inception_comparison}
\end{figure}
\paragraph{Experiments.}While DDPM is equivalent to DDCM with $K=\infty$, the first question we address is whether DDCM maintains the synthesis capabilities of DDPM for relatively small $K$ values.
We compare the performance of DDPM with that of DDCM using $K\in\{2,4,8,16,64\}$ for sampling from pre-trained pixel and latent space models.
We compute the Fréchet Inception Distance (FID)~\citep{heusel2017gans} to evaluate the generation performance.
In \Cref{app:more_random_gens} we report additional metrics and provide qualitative comparisons.
For pixel space generation, we use a class-conditional DDM trained on ImageNet $256\times256$~\citep{deng2009imagenet,dhariwal2021diffusion}, and apply classifier guidance (CG)~\citep{dhariwal2021diffusion} with unit scale.
We use the 50k validation set of ImageNet as the reference dataset, and sample 10k class labels randomly to generate the images.
For latent space, we use Stable Diffusion (SD) 2.1~\citep{rombach2022high} trained on $768\times768$ images and apply classifier-free guidance (CFG) with scale 3 (equivalent to $w=2$ in~\citep{ho2021classifier}).
As the reference dataset, we randomly select 10k images from MS-COCO~\citep{lin2014microsoft,chen2015microsoftcococaptionsdata} along with one caption per image, and use those captions as prompts for sampling.

As shown in~\Cref{fig:generation_fid_inception_comparison}, DDCM achieves similar FID scores to DDPM at $K=64$, suggesting that the Gaussian representation space of DDPM is redundant.
In the next sections we leverage our new representation space to solve a variety of tasks, including image compression and compressed image restoration.

\section{Image Compression with DDCM}\label{section:compression}
\begin{figure*}[t]
    \centering
    \includegraphics[width=1\textwidth]{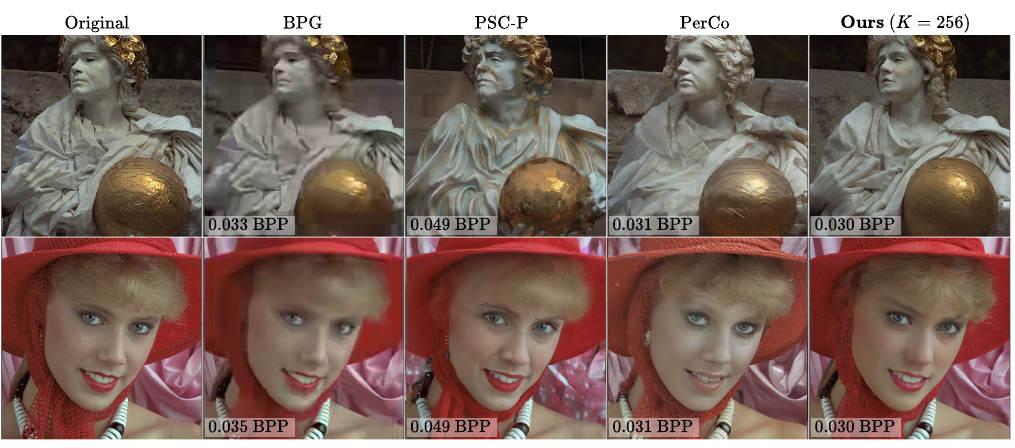}
    \caption{\textbf{Qualitative image compression results.} The presented images are taken from the Kodak24 ($512\times 512$) dataset.
    Our codec produces highly realistic outputs, while maintaining better fidelity to the original images compared to previous methods.
    }
    \label{fig:compression_examples}
\end{figure*}
\paragraph{Method.}
Since sampling with DDCM yields compact bit-stream representations, a natural endeavor is to harness DDCM for compressing real images.
In particular, to compress an image $\rvx_{0}$, we leverage the predicted $\hat{\rvx}_{0|i}$ (\Cref{eq:x0eq}) at each timestep $i$ and compute the residual error from the target image, $\rvx_0-\hat{\rvx}_{0|i}$.
Then, we guide the sampling process towards $\rvx_0$ by selecting the codebook entry that maximizes the inner product with this residual,
\begin{align}\label{eq:compression_rule} 
    k_i = \argmax_{k\in\{1,\hdots,K\}} \langle \gC_i(k), \rvx_0-\hat{\rvx}_{0|i}\rangle,
\end{align}
where the size of the first codebook $\gC_{T+1}$ is $K=1$. 
This process is depicted as the compression branch in \Cref{fig:overview}, where the resulting set of chosen indices $\{k_i\}_{i=2}^{T+1}$ is the compressed bit-stream representation of the given image.
Section~\ref{sec:compressed_conditional_generation} sheds more light on this choice of the noise selection from the perspective of score-based generative models~\citep{song2020score}. 
As in \Cref{sec:method}, decompression follows standard DDCM sampling \cref{eq:DDCM_sampling}, re-selecting the stored indices instead of picking them randomly.
When using latent space DDMs (e.g., SD), we 
first encode $\rvx_0$ into the latent domain, perform all subsequent operations in this domain, and decode the result with the decoder.

The bit rate of this approach is determined by the size of the codebooks $K$, and the number of sampling timesteps $T$. Specifically, the bit-stream length is given by \smash{$(T-1)\log_{2}(K)$}. Therefore, the bit rate can be reduced by simply decreasing the number of codebooks, or by using a smaller number of timesteps at generation, e.g., by skipping every other step, or by using specific timestep intervals (see \Cref{app:range_t}).
In the approach described so far, the length of the bitstream increases logarithmically with $K$, making it computationally demanding to increase the bit rate.
For instance, even for $K=8192$, $T=1000$ and $768\times 768$ images our BPP is approximately 0.022.
Thus, to produce higher bit rates, we propose to \emph{refine} the noise selected at timestep $i$ by employing matching pursuit (MP)~\citep{mallat1993matching}.
Specifically, at each step $i$, we construct the chosen noise as a convex combination of $M$ elements from 
$\gC_i$, gathered in a greedy fashion to best correlate with the guiding residual $\rvx_{0}-\hat{\rvx}_{0|i}$ (as in~\Cref{eq:compression_rule}). 
The resulting convex combination involves $M-1$ quantized scalar coefficients, chosen from a finite set of $C$ values taken from $[0,1]$.
Therefore, the resulting length of the bit-stream is given by $\smash{(T-1)(\log_{2}(K)M+C(M-1))}$, such that $M=1$ is similar to our standard compression scheme, and the length of the bit-stream increases linearly with $M$ and $C$.
We apply this algorithm when the absolute bits number crosses $(T-1)\cdot \log_2(2^{13})$.
Further details are available in \Cref{app:matching_pursuit}.

\paragraph{Experiments.}
We evaluate our compression method on Kodak24~\cite{franzen1999kodak}, DIV2K~\citep{agustsson2017ntire}, ImageNet 1K $256\times 256$~\citep{deng2009imagenet,pan2020dgp}, and CLIC2020~\citep{CLIC2020}.
For all datasets but ImageNet, we center crop and resize all images to $512\times512$.
We compare to numerous competing methods, both non-neural and neural, and both zero-shot, fine-tuning based, and training based approaches.
For the ImageNet dataset, we use the unconditional pixel space ImageNet $256\times 256$ model of \citet{dhariwal2021diffusion}, and compare our results to BPG~\citep{bpg}, HiFiC~\citep{mentzer2020high}, IPIC~\citep{xu2024idempotence}, and two  PSC~\citep{elata2024zero} configurations, distortion-oriented (PSC-D) and perception-oriented (PSC-P).
For all other datasets, we use SD 2.1 $512\times512$~\citep{rombach2022high} and compare to BPG, PSC-D, PSC-P, ILLM~\citep{muckley2023improving}, PerCo (SD)~\citep{korber2024perco, careil2023towards}, and twoCRDR~\citep{iwai2024controlling} configurations, distortion-oriented (CRDR-R) and perception-oriented (CRDR-R).
PSC shares the same pre-trained model as ours, while PerCo (SD) requires additional fine-tuning.
For our method, we apply SD 2.1 unconditionally, as we saw no improvement by adding prompts (see further details in \Cref{app:text_effect}).
We assess our method for several options of $T$, $K$, $M$, and $C$ to control the bit rate.
See further details in \Cref{app:compression_details}.
We evaluate distortion with PSNR and LPIPS~\citep{zhang2018perceptual} and perceptual quality with FID~\citep{bińkowski2018demystifying}.
For ImageNet, FID is computed against the entire 50k $256\times 256$ validation set.
For the smaller datasets we follow \citet{mentzer2020high} and compute the FID over extracted image patches. 
Specifically, for DIV2K and CLIC2020 we extract $128\times 128$ sized patches, and for Kodak we use $64\times64$.

\begin{figure*}[t]
    \centering
    \includegraphics[width=\linewidth]{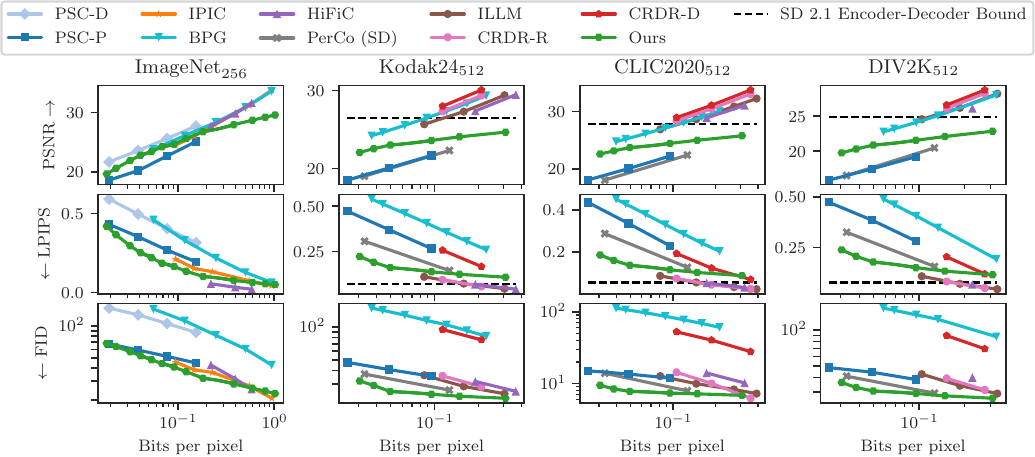}
    \caption{\textbf{Compression quantitative evaluation.}
    We compare the perceptual quality (FID) and distortion (PSNR, LPIPS) achieved at different BPPs. 
    The image sizes of each dataset is denoted next to its name.
    Our method produces the best perceptual quality at most BPPs.
    Importantly, this is while we also attain lower distortion compared to perceptually-oriented methods (e.g., PSC-P and PerCo (SD)). 
    For the three rightmost datasets, note that our approach, PSC-P, and PerCo (SD) use the latent space Stable Diffusion 2.1 model, so its encoder-decoder imposes a distortion bound.
    Thus, we report the distortion attained by simply passing the images through this encoder-decoder (dashed line).
    }
    \label{fig:compression_graphs}
\end{figure*}

As shown on the rate-distortion and rate-perception planes in \Cref{fig:compression_graphs}, our compression scheme dominates previous methods on the rate-perception-distortion tradeoff~\citep{pmlr-v97-blau19a} for lower bit rates, surpassing both the perceptual quality (FID) and distortion (PSNR and LPIPS) of previous methods.
For instance, our FID scores are lower than those of all other methods at around 0.1 BPP, while, for the same BPP, our distortion performance is better than the perceptually-oriented methods (e.g., PerCo, PSC-P, and IPIC).
However, our method under-performs at the highest bit rates, especially when using SD. 
When using a latent DDM such as SD, we aim to compress the latent encoding of a given image, rather than the image itself. Since encoding and decoding an image typically leads to a distorted reconstruction, the distortion of the compression method is bounded by that of the encoder-decoder. This leads to a performance ceiling for any latent-space-based compression method~\citep{korber2024perco, elata2024zero}.
We therefore include in \Cref{fig:compression_graphs} the ``SD 2.1 Encoder-Decoder bound'', which corresponds to the distortion resulting from encoding and decoding the original images using the VAE of SD 2.1, without any additional compression.
The qualitative comparisons in \Cref{fig:compression_examples} further demonstrate our superior perceptual quality, where, even for extreme bit rates, our method maintains similarity to the original images in fine details.
See \Cref{app:compression_details} for more details and results.

\section{Compressed Conditional Generation}\label{sec:compressed_conditional_generation}

We showed that DDCM can be used as an image codec by following a simple index selection rule, guiding the generated image towards a target one.
Here, we generalize this scheme to any \emph{conditional} generation task, considering the more broad framework of \emph{compressing} conditionally generated samples.
This is a particularly valuable framework in scenarios where the input condition $\rvy$ is bit rate intensive, e.g., where $\rvy$ is a degraded image and the goal is to produce a \emph{compressed} high-quality reconstruction of it.
To the best of our knowledge, this task, which we name \emph{compressed} conditional generation, has only been thoroughly explored for text input conditions~\citep{bulla2023}.

A naive solution to this task could be to simply compress the outputs of any existing conditional generation scheme.
Here we propose a novel end-to-end solution that generates the outputs \emph{directly} in their compressed form.
Importantly, note that our approach in \Cref{sec:method} requires the condition $\rvy$ for decompressing the bit-stream.
While this is not a stringent requirement when the condition is lightweight (e.g., a text prompt), this approach is less suitable when storage of the condition signal itself requires a long bit-stream. 
The solutions we propose in this section enable decoding the bit-stream without access to $\rvy$.

\paragraph{Compressed Conditional Generation with DDCM.}We propose generating a conditional sample by choosing the indices $k_{i}$ in \Cref{eq:DDCM_sampling} via
\begin{align}
k_{i}=\argmin_{k\in\{1,\hdots,K\}}\gL(\rvy,\rvx_{i},\gC_{i},k),\label{eq:k_choose_conditional}
\end{align}
instead of picking them randomly.
Here, $\gL(\rvy,\rvx_{i},\gC_{i},k)$ can be any loss function that attains a lower value when $\gC_{i}(k)$ 
directs the generative process towards an image that matches $\rvy$.
For example, for the loss
\begin{align}
\gL_{\text{P}}(\rvy,\rvx_{i},\gC_{i},k)=\norm{\gC_{i}(k)-\sigma_{i}\nabla_{\rvx_{i}}\log{p_{i}(\rvy|\rvx_{i})}}^{2}\label{eq:l_score}
\end{align}
we obtain the following result (see proof in~\cref{appendix:cond_compression}):
\begin{proposition}\label{prob:ode_convergence}
Suppose that image samples are generated via process~\cref{eq:DDCM_sampling}, and the indices $k_{i}$ are chosen according to~\Cref{eq:k_choose_conditional} with $\gL=\gL_{\textnormal{P}}$.
Then, when $K\rightarrow\infty$, such a generative process becomes a discretization of a probability flow ODE over the posterior distribution $p_{0}(\rvx_{0}|\rvy)$.
\end{proposition}
In other words,~\Cref{prob:ode_convergence} implies that for the loss $\gL_{\text{P}}$, increasing $K$ leads to more accurate sampling from the posterior $p_{0}(\rvx_{0}|\rvy)$, though this results in longer bit-streams.
Thus, as long as we have access to $\nabla_{\rvx_{i}}\log{p_{i}(\rvy|\rvx_{i})}$ (or an  approximation of it) $\gL_{\text{P}}$ may serve as a sensible option for solving a compressed conditional generation task with DDCM.
Interestingly, we show in~\Cref{appendix:compression_private_case} that our compression scheme from~\Cref{section:compression} is a special case of the proposed compressed conditional generation method, with $\rvy=\rvx_{0}$ and $\gL=\gL_{\text{P}}$.

\subsection{Compressed Posterior Sampling for Image Restoration}\label{sec:zero-shot-restoration}

\begin{figure*}[t]
    \centering
    \includegraphics[width=1.0\textwidth]{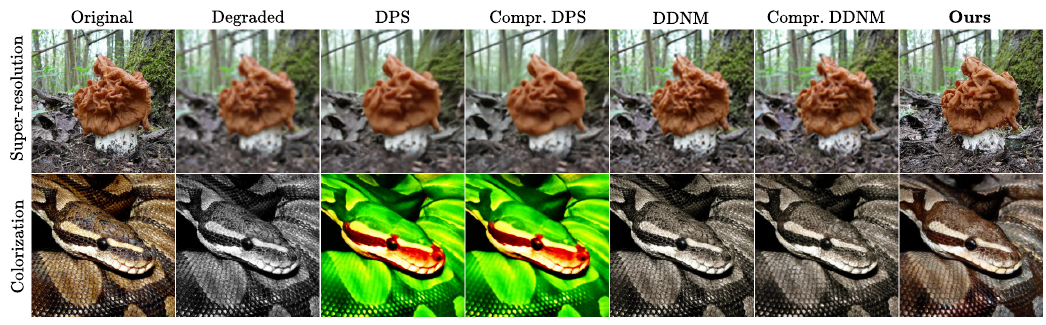}
    \includegraphics{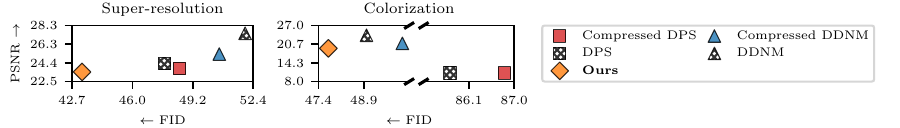}
    \caption{\textbf{Comparison of zero-shot posterior sampling image restoration methods.} Our approach achieves better perceptual quality compared to previous methods, while maintaining competitive PSNR and automatically producing compressed bit-stream representations for each restored image.}
    \label{fig:zero-shot-posterior-sampling-qualitative}
\end{figure*}

Our compressed conditional sampling approach can be utilized for solving inverse problems via posterior sampling.
Specifically, we consider inverse problems of the form $\rvy=\mA\rvx_{0}$, where $\mA$ is some linear operator.
We restrict our attention to \emph{unconditional} diffusion models and solve the problems in a ``zero-shot'' manner (similarly to \citet{kawar2022denoising,chung2023diffusion,wang2023zeroshot}). 
To generate conditional samples, we propose using the loss
\begin{align}
\gL(\rvy,\rvx_{i},\gC_{i},k)=\norm{\rvy-\mA(\vmu_{i}(\rvx_{i})+\sigma_{i}\gC_{i}(k))}^{2}.\label{eq:l_posterior_sampling}
\end{align}
Note that~\Cref{eq:l_posterior_sampling} attains a lower value when $\sigma_{i}\mA\gC_{i}(k)$ points in the direction that perturbs $\mA\vmu_{i}(\rvx_{i})$ towards $\rvy$.
Thus, our conditional generative process aims to produce a reconstruction $\hat{\rvx}$ that satisfies $\mA\hat{\rvx}\approx\rvy$, implying that we approximate posterior sampling~\citep{pmlr-v202-ohayon23a}.
Notably, when assuming that $p_{i}(\rvy|\rvx_{i})$ is a multivariate normal distribution centered around $\mA\rvx_{i}$ (as in~\citep{jalal2021posterior}), the chosen codebook noise $\gC_{i}(k_{i})$ approximates the gradient $\nabla_{\rvx_{i}}\log{p_{i}(\rvy|\rvx_{i})}$ and~\Cref{eq:l_posterior_sampling} becomes a proxy of~\Cref{eq:l_score}.

Following~\citep{chung2023diffusion,wang2023zeroshot}, we implement our method using the unconditional ImageNet $256\times 256$ DDM trained by \citet{dhariwal2021diffusion}.
We fix $K=4096$ for all codebooks, resulting in a compressed bit-stream of approximately $0.183$ BPP for each generated image.
We compare our method with DPS~\citep{chung2023diffusion} and DDNM~\citep{wang2023zeroshot} on two noiseless tasks: colorization and $4\times $ super-resolution (using the bicubic kernel). We evaluate these methods using their official implementations and the same DDM.
We additionally compress the outputs of DPS and DDNM to assess whether such a naive approach would yield better results.
To do so, we adopt our proposed compression scheme (from \Cref{section:compression}), employing the same unconditional ImageNet DDM and using $K=4096$ noises per codebook.

Qualitative and quantitative results are reported in \Cref{fig:zero-shot-posterior-sampling-qualitative}.
As expected, due to the rate-perception-distortion tradeoff~\citep{pmlr-v97-blau19a}, we observe that compressing the outputs of DPS and DDNM harms either their perceptual quality (FID), or their distortion (PSNR), or both.
This is while our method achieves superior perceptual quality compared to both DPS and DDNM, including their compressed versions.
While our method achieves slightly worse PSNR, this is expected due to the perception-distortion trade-off~\citep{Blau_2018_CVPR}.
See \Cref{appendix:zero-shot} for more details.

\subsection{Compressed Real-World Face Image Restoration}\label{sec:bfr}
\begin{figure*}[t]
    \centering
    \includegraphics[width=1\textwidth]{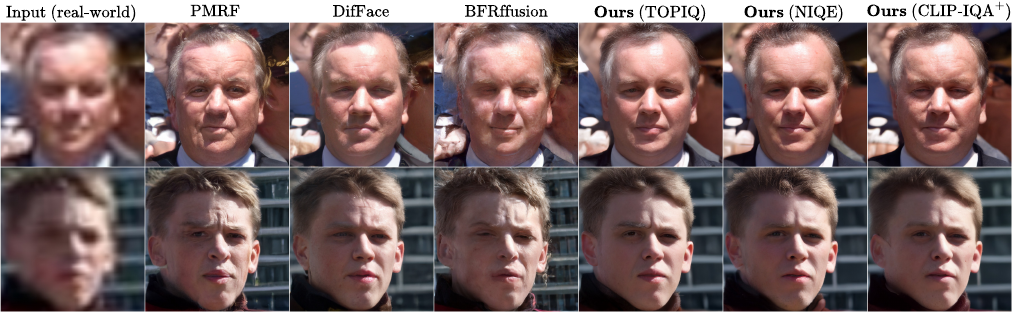}
\includegraphics[width=1\textwidth]{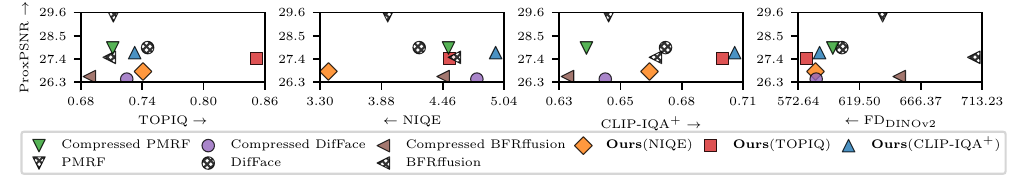}
    \caption{\textbf{Comparing real-world face image restoration methods on the WIDER-Test dataset}. We successfully optimize the NR-IQA measures and produce appealing output perceptual quality with less artifacts compared to previous methods.}
    \label{fig:real-world-wider-visual}
\end{figure*}
Real-world face image restoration is the practical task of restoring any degraded face image, without any knowledge of the corruption process it has gone through~\citep{wang2021gfpgan,vqfr,wang2022restoreformer,zhou2022codeformer,wang2023restoreformer++,2023diffbir,difface,bfrfussion,pmrf}.
We propose a novel method capable of optimizing any no-reference image quality assessment (NR-IQA) measure at test time (e.g., NIQE~\citep{niqe}), without relying on gradients.

Specifically, at each timestep $i$, we start by picking two indices -- one that promotes high perceptual quality, $k_{i,P}$, and another that promotes low distortion, $k_{i,D}$.
Then, we choose between $k_{i,P}$ and $k_{i,D}$ the index that better optimizes a desired balance of the perception-distortion tradeoff~\citep{Blau_2018_CVPR}.
Formally, letting $\vr(\rvy)\approx\mathbb{E}[\rvx_{0}|\rvy]$ denote the approximate Minimum Mean-Squared-Error (MMSE) estimator of this task, we pick $k_{i,D}$ via
\begin{align}
    k_{i,D}=\argmax_{k\in\{1,\hdots,K\}}\langle\gC_{i}(k),\vr(\rvy)-\hat{\rvx}_{0|i}\rangle.\label{eq:first_index_blind_face_restoration}
\end{align}
Note that this index selection rule is similar to that of our standard compression, replacing $\smash{\rvx_{0}}$ in \Cref{eq:compression_rule} with $\smash{\vr(\rvy)}$.
This choice of indices in DDCM would lead to a reconstructed estimate of the MMSE solution $\vy(\rvy)$, yielding blurry results with low distortion~\citep{Blau_2018_CVPR}.
In contrast, \emph{randomly} picking a sequence of indices in DDCM would produce a high quality sample from the data distribution $p_{0}$.
Therefore, we randomly choose $\smash{k_{i,P}\sim \text{Unif}(\{1,\hdots,K\})}$.
Then, we use the DDM and compute $\smash{\hat{\rvx}_{0|i-1}}$ for each index $\smash{k\in\{k_{i,D},k_{i,P}\}}$ separately, denoting each result accordingly by $\hat{\rvx}_{0|i-1}^{(k)}$.
The final index is picked to optimize the perception-distortion tradeoff via
\begin{align}
    k_{i}\!=\!\argmin_{k\in\{k_{i,D},k_{i,P}\}}\!\text{MSE}\!\left(\!\vr(\rvy),\hat{\rvx}_{0|i-1}^{(k)}\!\right)\!+\!\lambda Q\!\left(\!\hat{\rvx}_{0|i-1}^{(k)}\!\right)\!,\label{eq:optimize-perception-distortion-indices}
\end{align}
where $Q(\cdot)$ can be \emph{any} NR-IQA measure, even a non-differentiable one.
In \Cref{app:dmax-ot} we explain our choice to set $\vr(\rvy)$ as an MMSE estimator.

We assess our approach choosing $\vr(\rvy)$ as the FFHQ~\citep{stylegan} $512\times 512$ approximate MMSE model trained by \citet{difface}.
We set $\lambda=1$ and optimize three different $Q(\cdot)$ measures: NIQE, $\text{CLIP-IQA}^{+}$~\citep{Wang_Chan_Loy_2023}, and TOPIQ~\citep{chen2024topiq} adapted for face images by PyIQA~\citep{pyiqa}.
We utilize the FFHQ $512\times512$ DDM of \citet{difface} with $T=1000$ sampling steps and $K=4096$ for all codebooks.
We compare our approach against the state-of-the-art methods PMRF~\citep{pmrf}, DifFace~\citep{difface}, and BFRffusion~\citep{bfrfussion}, using the standard evaluation datasets CelebA-Test~\citep{karras2018progressive,wang2021gfpgan}, LFW-Test~\citep{lfw-original}, WebPhoto-Test~\citep{wang2021gfpgan}, and WIDER-Test~\citep{zhou2022codeformer}.
We use PSNR to measure the distortion of the outputs produced for the CelebA-Test dataset, and measure the ProxPSNR~\citep{pmrf,man2025proxiesdistortionconsistencyapplications} for the other datasets, which lack the clean original images.
Perceptual quality is measured by NIQE, $\text{CLIP-IQA}^{+}$, TOPIQ-FACE, and additionally $\text{FD}_{\text{DINOv2}}$~\citep{stein2023exposing} to assess our generalization performance to a common quality measure which we do not directly optimize.
Finally, as in \Cref{sec:zero-shot-restoration}, we \emph{compress} each evaluated method using our standard compression scheme, adopting the same FFHQ DDM with $K=4096$ and $T=1000$.

The results for the WIDER-Test dataset are reported in \Cref{fig:real-world-wider-visual} (see \Cref{app:dmax-ot} for the other datasets).
Our approach clearly optimizes each quality measure effectively and generalizes well according to the $\text{FD}_{\text{DINOv2}}$ scores.
This is also confirmed visually, where all of our solutions produce high-quality images with less artifacts compared to previous methods.
While our approach shows slightly worse distortion, this is once again expected due to the perception-distortion tradeoff~\citep{Blau_2018_CVPR}.

\section{Discussion}\label{sec:discussion}

We introduced DDCM, a novel generative approach for DDMs that produces high-quality image samples jointly with their lossless compressed bit-stream representations.
We found that DDCM achieves comparable generative performance to DDPM, even when the codebooks are extremely small.
We leveraged DDCM to solve several compressed image generation tasks, including image compression and compressed restoration, where we achieved state-of-the-art results.
Besides image restoration, our compressed conditional generation framework can be used for any type of diffusion guidance, e.g., for text-conditional generation.
We demonstrate this option in~\Cref{appendix:classifier-guidance,appendix:classifier-free-guidance}, introducing new classifier-based and classier-free guidance methods that do not use $\rvy$ for decompression.
Moreover, we present in \Cref{app:editing} preliminary results for compressed image editing using DDCM, by decompressing an image using a desired edit text prompt.

While our empirical results are encouraging, our work does not explain theoretically why DDCM sampling and our simple index selection strategies work so effectively.
We encourage future works to investigate the principles behind the success of our methods.
Moreover, when operating in latent space, our codec's performance is bounded by the underlying VAE, particularly at higher bit rates.
Our results at higher bit rates could also be improved through better approaches than our current matching pursuit inspired solution.
Additionally, our compression efficiency could be improved through entropy coding of the selected indices, potentially reducing bit rates without sacrificing quality.
Lastly, all DDCM-based solutions can be improved further by optimizing the codebooks, e.g., through dictionary learning.

While DDCM produces compressed image representations in the form of indices, it is worth noting that the codebook entries can be interpreted as discrete image tokens, which are chosen autoregressively.
This implies that DDCM can be interpreted as an image tokenization method, forming a ``language'' for images.
Such a perspective opens the door to intriguing possibilities, such as using the tokenized image representation as a natural text-conditioning mechanism for unconditional diffusion models. For instance, one could train a transformer to predict the sequence of codebook indices corresponding to an image given its text description.
Such a transformer can then be used to generate text-conditional images, by predicting the corresponding sequence of DDCM codebook indices.

To conclude, our work demonstrates promising results across numerous tasks while providing opportunities for both theoretical analysis and practical improvements.


\section*{Acknowledgements}
This research was partially supported by the Israel Science Foundation (ISF) under Grants 2318/22, 951/24 and 409/24, and by the Council for Higher Education – Planning and Budgeting Committee.
We thank Noam Elata and Matan Kleiner for assisting with our figures and experiments.

\section*{Impact Statement}
This paper presents work whose goal is to advance the field of Machine Learning.
There are many potential societal consequences of our work, none which we feel must be specifically highlighted here.

\bibliographystyle{icml2025}
\bibliography{citations}

\newpage
\appendix
\onecolumn

\section{Additional DDCM Evaluation}\label{app:more_random_gens}

\begin{figure}[H]
    \centering
    \includegraphics[width=0.5\linewidth]{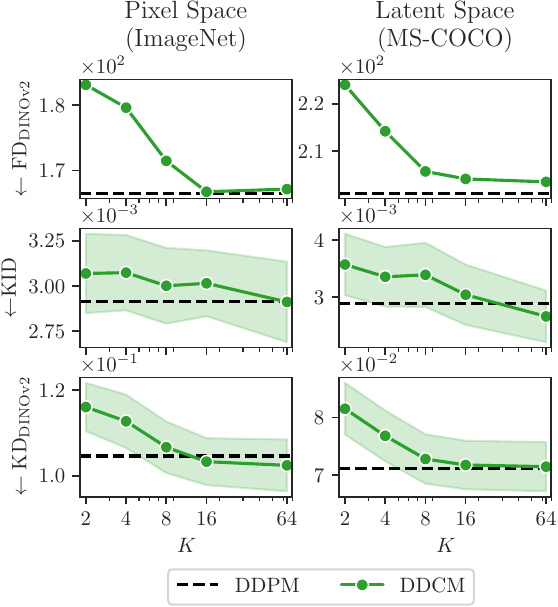}
    \caption{\textbf{Comparing DDPM with DDCM for different codebook sizes $K$.} 
    As in \Cref{sec:method}, DDPM and DDCM with $K=64$ (sometimes even $K=16$) achieve similar generative performance, suggesting that the continuous representation space of DDPM (DDCM with $K =\infty$) is highly redundant.
    We use a class-conditional ImageNet model ($256 \times 256$) for pixel space, and the text-conditional SD 2.1 model ($768 \times 768$) for latent space. The $K$ axis is in log-scale.
    }
    \label{fig:generation_other_metrics}
\end{figure}

\begin{figure*}[t]
    \centering
    \includegraphics[width=\linewidth]{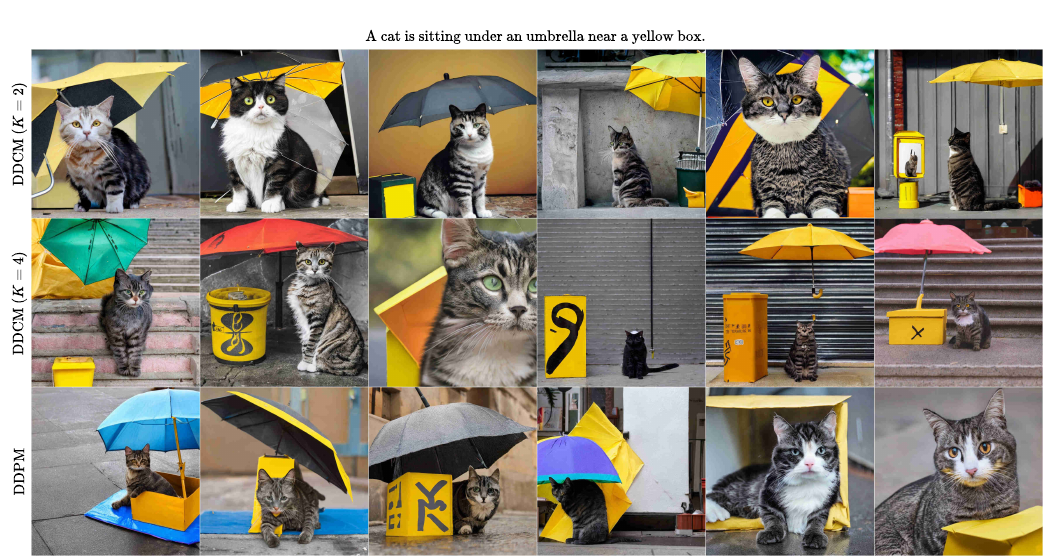}\vspace{0.2cm}
    \includegraphics[width=\linewidth]{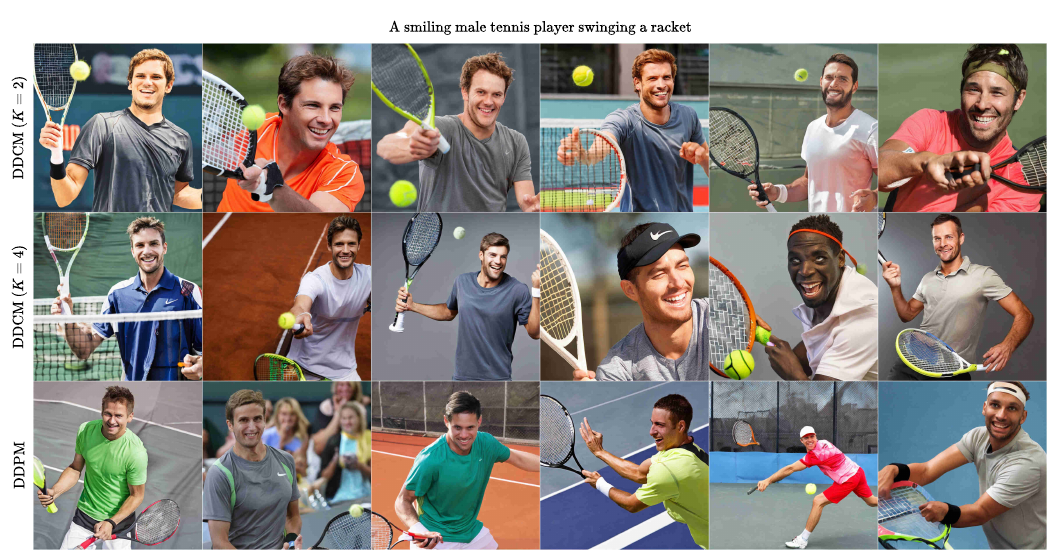}\vspace{0.2cm}
    \caption{\textbf{Qualitative comparison of sample quality and diversity between DDCM and DDPM.}
    We generate multiple samples for each prompt, using the $768\times768$ SD 2.1 model.}
    \label{fig:generation_samples_latent_app3}
\end{figure*}

\begin{figure*}[t]
    \centering
    \includegraphics[width=1\textwidth]{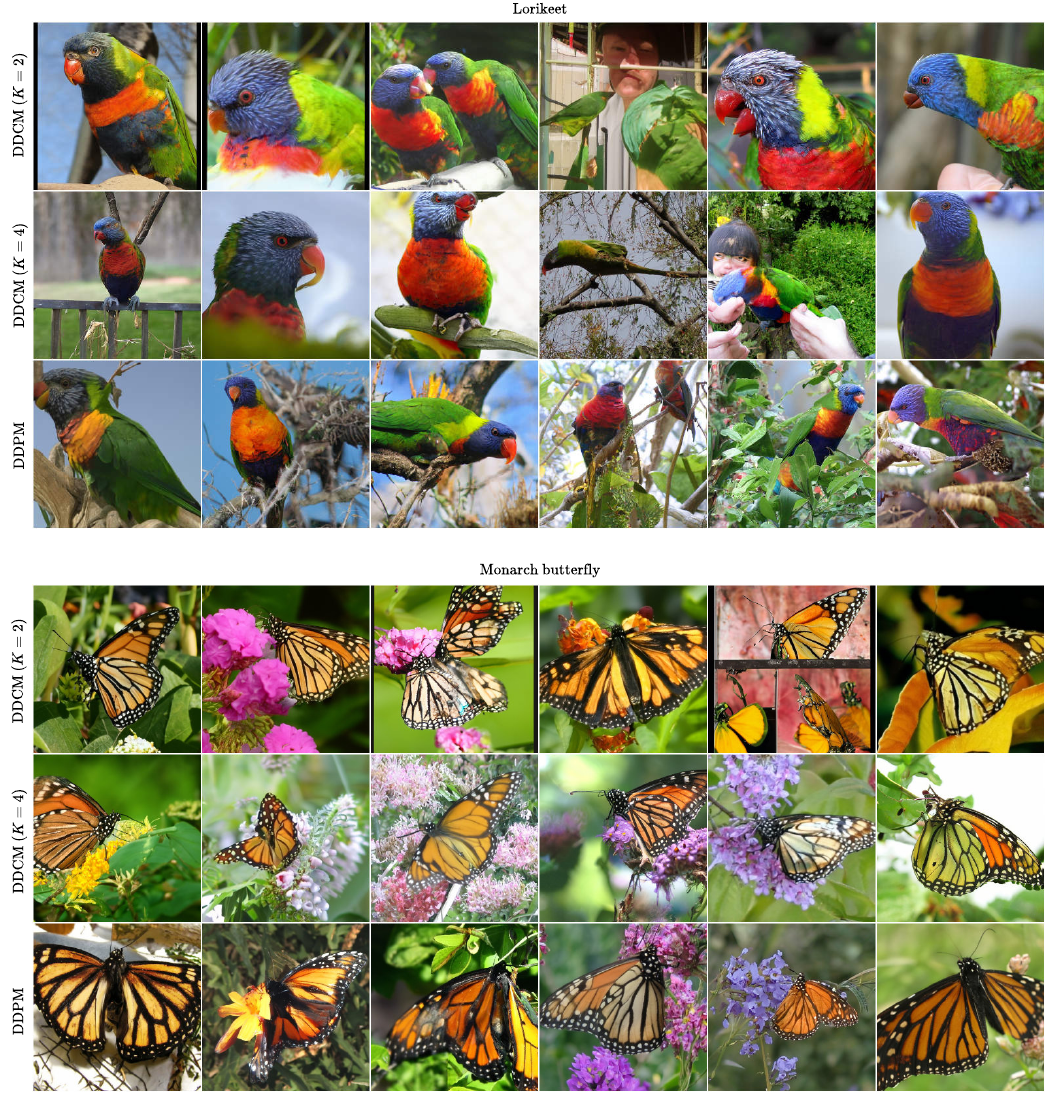}
    \caption{\textbf{Qualitative comparison of sample quality and diversity between DDCM and DDPM.}
    We generate multiple samples for each class, using the conditional $256\times 256$ ImageNet model.}
    \label{fig:generation_samples_pixel_app3}
\end{figure*}

Figure~\ref{fig:generation_other_metrics} provides additional quantitative comparisons between DDPM and DDCM, using different $K$ values.
Specifically, we compute the Kernel Inception Distance (KID)~\citep{bińkowski2018demystifying}, as well as the Fréchet Distance and Kernel Distance evaluated in the feature space of DINOv2~\citep{stein2023exposing,oquab2024dinov}.
These quantitative results remain consistent with the ones presented in \Cref{sec:method}, showing that DDCM with small $K$ values is comparable with DDPM.
Figures~\ref{fig:generation_samples_latent_app3} and \ref{fig:generation_samples_pixel_app3} show numerous outputs from both DDPM and DDCM with small values of $K$, demonstrating the sample quality and diversity produced by the latter for such $K$ values.
We use Torch Fidelity~\citep{obukhov2020torchfidelity} to compute the perceptual quality measures.


\clearpage

\section{Image Compression Supplementary}\label{app:compression_details}

We compute all distortion and perceptual quality measures using \href{https://github.com/Lightning-AI/torchmetrics}{Torch Metrics} (which relies on Torch Fidelity~\citep{obukhov2020torchfidelity}).

\subsection{Experiment Configurations}\label{app:image-compression-experiments-configuration}
We specify here the different configurations used for the compression experiments in \Cref{section:compression}.
\begin{itemize}[parsep=2pt,itemsep=2pt]
    \item In our $256\times256$ experiments we use the \texttt{256x256\_diffusion\_uncond.pt} checkpoint from the \href{https://github.com/openai/guided-diffusion}{official GitHub repository}.
    In our $768\times768$ and $512\times512$ experiments we use Stable Diffusion 2.1 with the \href{https://huggingface.co/stabilityai/stable-diffusion-2-1}{\texttt{stabilityai/stable-diffusion-2-1}} and \href{https://huggingface.co/stabilityai/stable-diffusion-2-1-base}{\texttt{stabilityai/stable-diffusion-2-1-base}} official checkpoints from Hugging Face, respectively.
    The different $K$, $M$, $C$ and $T$ values for each of the $512\times512$ and the $768\times768$ experiments plotted in \Cref{fig:compression_graphs,fig:compression_graphs_768} are summarized in \Cref{tab:our_tmkc_configs}.

\begin{table}[h]
\centering
\caption{Image compression experiments configurations.}
\begin{tabular}{c|c|c|c|c|c}
\hline
Model & Image Resolution & $T$ & $K$ & $M$ & $C$ \\
\hline
\multirow{4}{*}{Pixel Space DDM} & \multirow{4}{*}{256$\times$256} & 1000 & 64, 128, 256, 4096 & 1 & - \\
& & 1000 & 2048 & 2, 3, 4, 5 & 3 \\
& & 500 & 128, 512 & 1 & - \\
& & 300 & 16, 32, 128, 512 & 1 & - \\
\hline
\multirow{6}{*}{Latent Space DDM} & \multirow{2}{*}{512$\times$512} & 1000 & 256, 1024, 8192 & 1 & - \\
& & 1000 & 2048 & 2, 3, 6 & 3 \\
\cline{2-6}
& \multirow{4}{*}{768$\times$768} & 1000 & 16, 32, 64, 256, 1024, 8192 & 1 & - \\
& & 1000 & 2048 & 2, 3, 6 & 3 \\
& & 500 & 16, 32, 64, 256, 1024, 8192 & 1 & - \\
& & Adapted 500 (\Cref{app:range_t}) & 16, 32, 64, 256, 1024, 8192 & 1 & - \\
\hline
\end{tabular}
\label{tab:our_tmkc_configs}
\end{table}


    
    \item For PSC-D and PSC-R we use the same pre-trained ImageNet $256\times256$ model as ours in the $256\times256$ experiments, and the same Stable Diffusion 2.1 model in the $512\times512$ experiments. We adopt the default hyper-parameters of the method as described by the authors~\citep{elata2024zero}, setting the number of measurements to $12\cdot2^i$ for $i=0,\ldots,8$.
    \item For IPIC we adopt the official implementation using the ELIC codec with five bit rates, combined with DPS sampling for decoding with $T=1000$ steps and $\zeta\in\{0.3, 0.6, 0.6, 1.2, 1.6\}$, as recommended by the authors.
    \item For BPG we considered quality factors $q\in\{51,50,48,46,42,40,38,36,34,32,30\}$. 
    \item For HiFiC we test the low, medium and high quality regimes, using the checkpoints available in the \href{https://github.com/Justin-Tan/high-fidelity-generative-compression}{official GitHub repository}.
    \item PerCo (SD) is tested using the three publicly available Stable Diffusion 2.1 fine-tuned checkpoints from their \href{https://github.com/Nikolai10/PerCo}{Official GitHub repository}, using the default hyper-parameters.
    \item For ILLM we use the MS-ILLM pre-trained models available in the \href{https://github.com/facebookresearch/NeuralCompression/tree/main/projects/illm}{official GitHub implementation}.
    For the $512\times512$ image size experiments we use \texttt{msillm\_quality\_X}, $\texttt{X}=2,3,4$. For the $768\times768$ image size experiments we use \texttt{msillm\_quality\_X}, $\texttt{X}=2,3$ and \texttt{msillm\_quality\_vloY}, $\texttt{Y}=1,2$.
    \item CRDR-D and CRDR-R are evaluated using quality factors of $\{0,1,2,3,4\}$, where CRDR-D uses $\beta=0$ and CRDR-R uses $\beta=3.84$, as recommended in the paper.
\end{itemize}

\subsection{Additional Evaluations}\label{app:compression_more_results}

\begin{figure*}[t]
    \centering
    \includegraphics[width=0.86\linewidth]{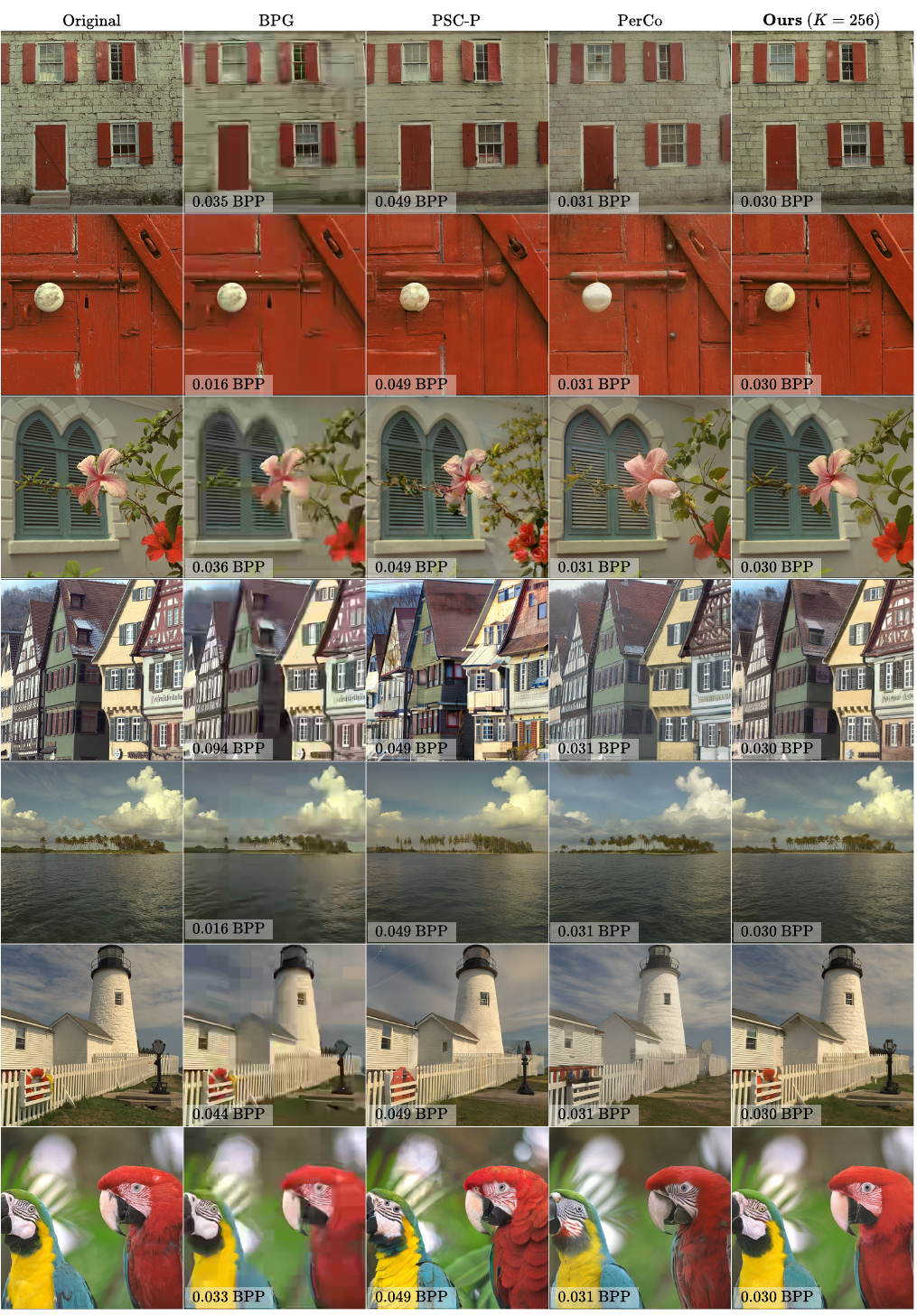}
    \caption{\textbf{Qualitative extreme image compression results.} The presented images are taken from the Kodak24 dataset, cropped to $512\times512$ pixels.
    Our compression scheme produces highly realistic decompressed outputs, while maintaining better fidelity to the original images compared to previous methods.
    }
    \label{fig:compression_examples_app}
\end{figure*}

\begin{figure*}[t]
    \centering
    \includegraphics[height=0.93\textheight]{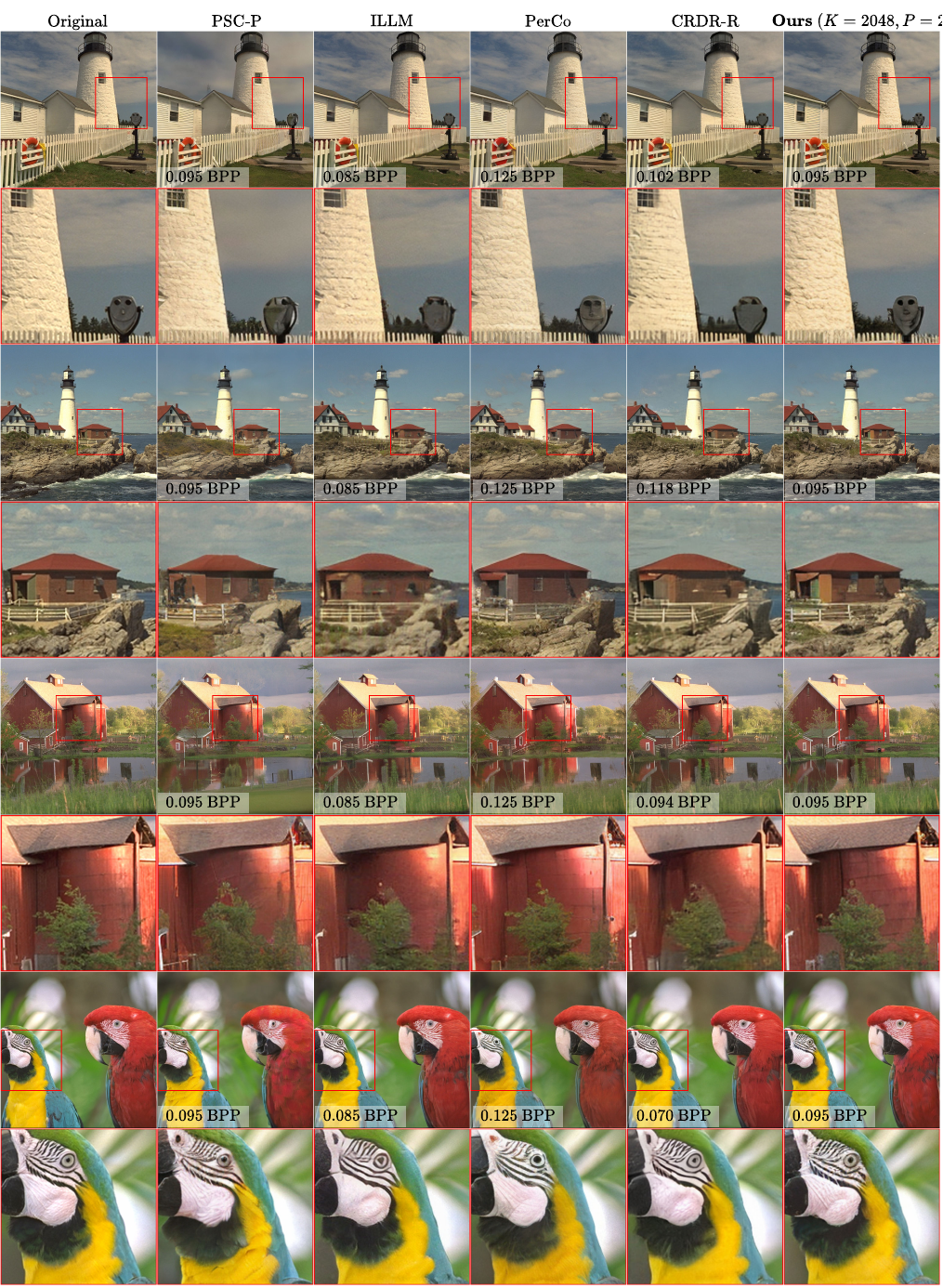}
    \caption{
    \textbf{Qualitative image compression results.} The presented images are taken from the Kodak24 dataset, cropped to $512\times512$ pixels.
    Our compression scheme produces highly realistic decompressed outputs, while maintaining better fidelity to the original images compared to previous methods.
    }\label{fig:compression_examples_app_medium}
\end{figure*}

\begin{figure*}[t]
    \centering
    \includegraphics[width=\linewidth]{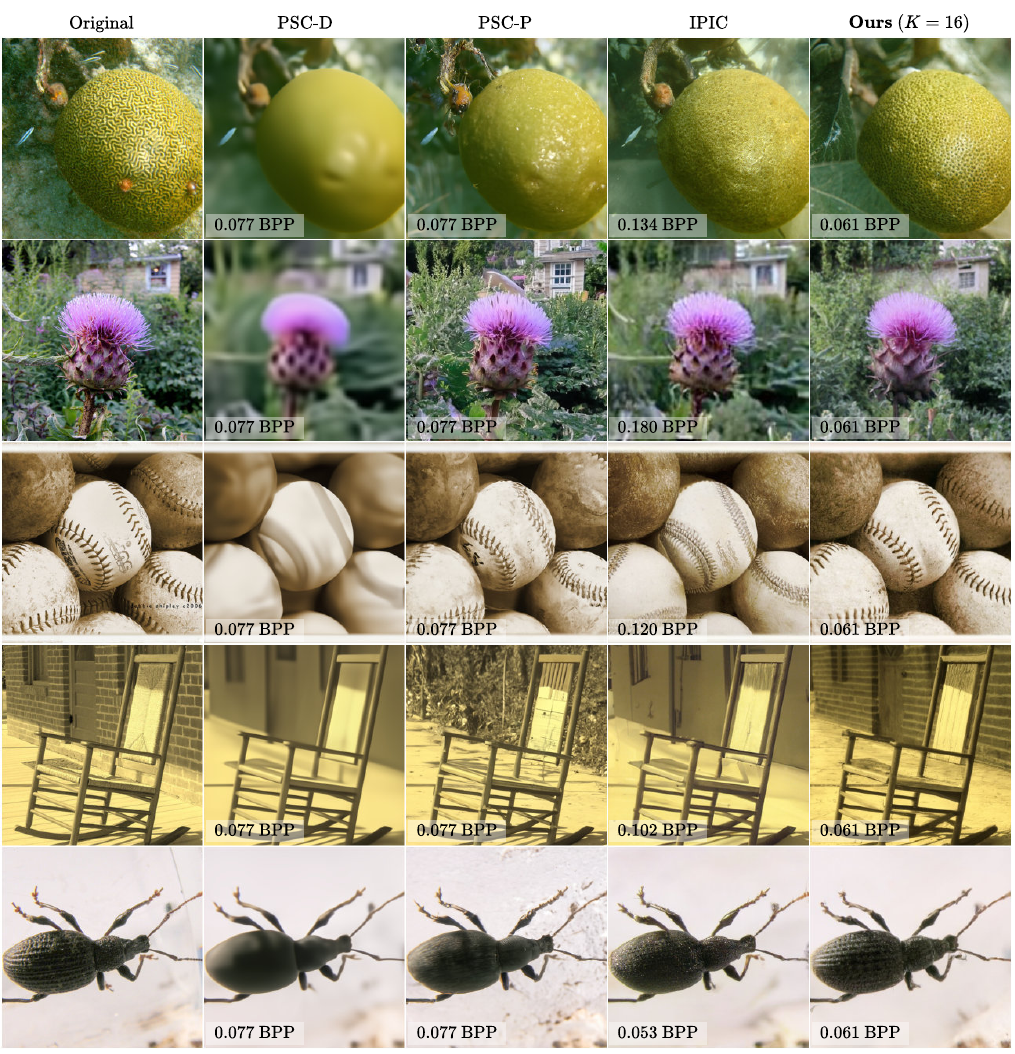}
    \caption{\textbf{Qualitative image compression results.} the presented images are taken from the ImageNet $256\times 256$ dataset.
    Compared to previous methods, our compression scheme produces higher perceptual quality and better fidelity to the original images.
    }
    \label{fig:pixel_space_comparison_low_bpp}
\end{figure*}

\begin{figure*}[t]
    \centering
    \includegraphics[width=\linewidth]{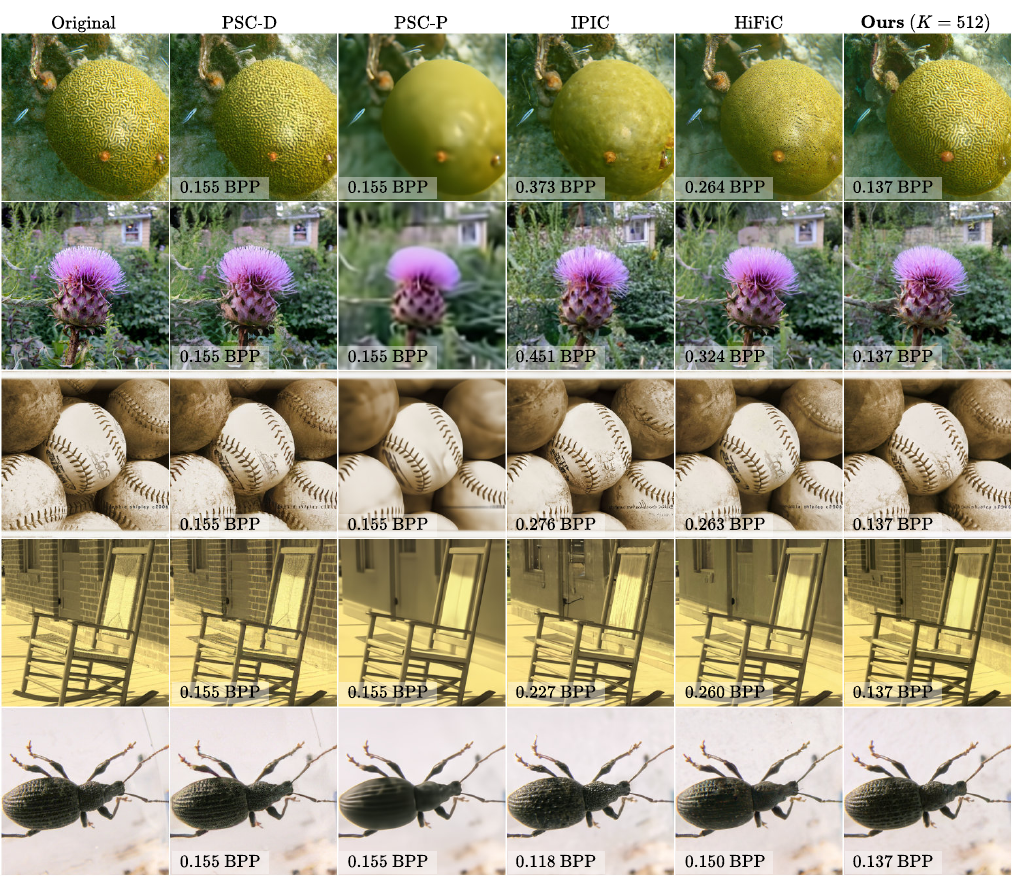}
    \caption{\textbf{Qualitative image compression results.} the presented images are taken from the ImageNet $256\times 256$ dataset.
    Compared to previous methods, our compression scheme produces higher perceptual quality and better fidelity to the original images.
    }
    \label{fig:pixel_space_comparison_mid_bpp}
\end{figure*}

In \Cref{fig:compression_examples_app,,fig:compression_examples_app_medium,,fig:pixel_space_comparison_low_bpp,,fig:pixel_space_comparison_mid_bpp} we provide additional qualitative comparisons on the Kodak24 ($512\times 512$) and ImageNet ($256\times 256$) datasets.
We additionally compare our method on images of size $768\times768$, and present the results in \Cref{fig:compression_graphs_768}.
Our method is implemented as before, while using a Stable Diffusion 2.1 model trained on the appropriate image size (see \Cref{app:image-compression-experiments-configuration}).

\subsection{Numerical Results}
\Cref{tab:numerical_compression_imagenet,,tab:numerical_compression_kodak,,tab:numerical_compression_clic,,tab:numerical_compression_div2k} include the numerical results that appear in \Cref{fig:compression_graphs}. Additionally, \Cref{tab:numerical_compression_clic_768,,tab:numerical_compression_div2k_768} include the numerical results for \Cref{fig:compression_graphs_768}.

\subsection{Decreasing the Bit Rate via Timestep Sub-Sampling}\label{app:range_t}

As mentioned in \Cref{section:compression}, decreasing the bit rate of our compression scheme can be accomplished in two ways.
The first option is to reduce $K$, which sets the number of bits required to represent each communicated codebook index.
The second option relates to the number of generation timesteps, which sets the total number of communicated indices. Specifically, DDMs trained for $T=1000$ steps can still be used to generate samples with fewer steps, by skipping alternating timesteps and modifying the variance in \Cref{eq:ddpm}. 
Thus, we leverage such timestep sub-sampling in DDCM to shorten the compressed bit-stream.
We find that pixel space DDMs yields good results with this approach, while the latent space models struggle to produce satisfying perceptual quality.

Thus, for latent space models we propose a slightly different timestep sub-sampling scheme. 
Specifically, we keep $T=1000$ sampling steps at inference and set different $K$ values for different subsets of timesteps.
We choose $K=1$ for a subset of $L$ sampling steps, and $K>1$ for the rest $T-L$ steps.
Thus, our compression scheme only optimizes $T-L$ steps and necessitates transmitting only $T-L$ indices. The rest $L$ indices correspond to codebooks that contain only one vector, and thus do not affect the bit rate.

We use $T=1000$, set the same codebook size $K>1$ for every timestep $i\in\{899\ldots,400\}$, and use $K=1$ for all other steps.
We compare our proposed method against the aforementioned naive timestep skipping approach with $T=500$ sampling steps and the same $K>1$, which attains the same bit rate as our proposed alternative.

\begin{figure*}[t]
    \centering
    \includegraphics[width=1\linewidth]{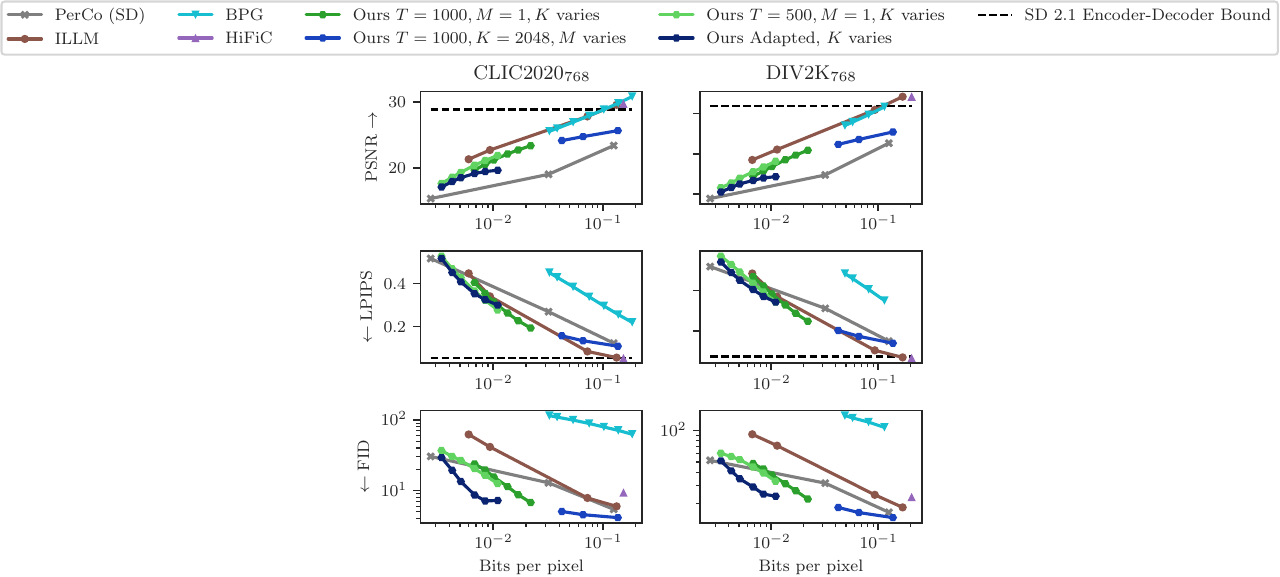}
    \caption{\textbf{Quantitative image compression results on $768\times768$ sized images.} At higher bit rates, our method achieves the lowest (best) FID scores in both datasets while maintaining better distortion metrics compared to PerCo (SD). At extremely low bit rates, while PerCo (SD) shows marginally better FID scores, our method attains superior PSNR performance.}
    \label{fig:compression_graphs_768}
\end{figure*}

Quantitative results are shown in \Cref{fig:compression_graphs_768}, where our timestep adapted method is denoted by \emph{Ours Adapted}.
Our adapted approach achieves better perceptual quality compared to the naive one, at the expense of a slightly hindered PSNR.

\subsection{Increasing the Bit Rate via Matching Pursuit}\label{app:matching_pursuit}
In \Cref{section:compression} we briefly explain how to achieve higher bit rates with our method, by refining each selected noise via a matching pursuit inspired solution.
Formally, at each timestep $i$ we iteratively refine the selected noise by linearly combining it with $M-1$ other noises from the codebook.
We start by picking the first noise index $k_{i}$ according to \Cref{eq:compression_rule}.
Then, we set $k_{i}^{(1)}=k_{i}$, $\gamma_{i}^{(1)}=1$, and $\tilde{\vz}_{i}^{(1)}=\gC_{i}(k_{i})$, and pick the next indices and coefficients $(m=1,\hdots,M-1)$ via
\begin{align}
    k_i^{(m+1)}, \gamma_i^{(m+1)} = \argmax_{k\in\{1,\hdots,K\},\;\gamma \in \Gamma} \left\langle \gamma\tilde{\vz}_{i}^{(m)}+\left(1-\gamma\right)\gC_{i}(k) ,\,\rvx_0-\hat{\rvx}_{0|i}\right\rangle.
\end{align}
The noise vector $\tilde{\vz}_{i}^{(m+1)}$ is then updated via
\begin{align}
&\tilde{\vz}_{i}^{(m+1)}\leftarrow\gamma_{i}^{(m+1)}\tilde{\vz}_{i}^{(m)}+\left(1-\gamma_{i}^{(m+1)}\right)\gC_{i}\left(k_{i}^{(m+1)}\right),\\
&\tilde{\vz}_{i}^{(m+1)}\leftarrow \frac{\tilde{\vz}_{i}^{(m+1)}}{\text{std}\left(\tilde{\vz}_{i}^{(m+1)}\right)},
\end{align}
where $\text{std}(\vz)$ is the empirical standard deviation of the vector $\vz$.
We use the resulting vector $\tilde{\vz}_{i}^{(M)}$ as the noise in \Cref{eq:DDCM_sampling} to produce the next $\rvx_{i-1}$, and repeat the above process iteratively.
Note that setting $M=1$ is equivalent to our standard compression scheme.

In our experiments, the set of coefficients $\Gamma$ is a subset of $(0,1]$, containing $C=|\Gamma|$ values that are evenly spaced in this range.
We pick $C=3$ and assess $M\in\{2,3,6\}$ for the latent space model experiments, and $M\in\{2,3,4,5\}$ for the pixel space model experiments.

\subsection{Assessing the Effectiveness of Text Prompts in Compression using Text-to-Image Latent Space DDMs}\label{app:text_effect}

Stable Diffusion 2.1 is a text-to-image generative model, which both PerCo (SD) and PSC leverage for their compression approach.
Specifically, both of these methods start by generating a textual caption for every target image using BLIP-2~\citep{li2023blip}, and feed the captions as prompts to the SD model.
In our case, we find that using such prompts hinders the compression quality.
Specifically, we follow the same automatic captioning procedure as in PerCo (SD) and PSC, using the \texttt{Salesforce/blip2-opt-2.7b-coco} \href{https://huggingface.co/Salesforce/blip2-opt-2.7b-coco}{checkpoint} of BLIP-2 from Hugging Face.
We then continue with our standard compression approach, where the denoiser is used with standard classifier-free guidance (CFG).
Note that using text prompts requires transmitting additional bits that serve as a compressed version of the text.
Specifically, we use BLIP-2 with a maximum of $L=32$ word tokens, each picked from a dictionary containing a total of 30,524 words.
Thus, at most $32\cdot\log_{2}(30524)\approx 480$ bits are added to the bit-stream in our method.

\begin{figure*}[t]
    \centering
    \includegraphics[width=0.9\linewidth]{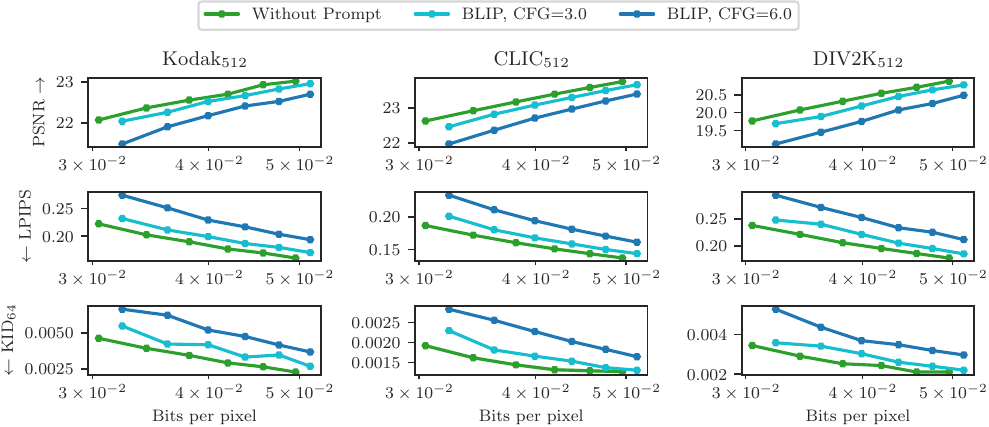}
    \caption{\textbf{Evaluating the effectiveness of using text prompts in image compression.}
    We evaluate our unconditional compression method with the text-conditional one, while using the text captions generated by BLIP-2.
    We find that using such text prompts hinders our compression results, both in terms of perceptual quality and distortion.}
    \label{fig:compression_graphs_blip}
\end{figure*}

We assess this text-conditional approach on the $512\times 512$ SD 2.1 DDM, using CFG scales of $3,6$.
We compare the performance of this conditional approach with that of the unconditional one we used in \Cref{section:compression}.
The results in \Cref{fig:compression_graphs_blip} show a disadvantage for using text-prompts for compression with our method.

\clearpage

\begin{table}[t]
    \centering
    \caption{Compression quantitative evaluations, for the ImageNet $256\times256$ dataset.}
    \begin{tabular}{lcccc}
\toprule
\textbf{Method} & \textbf{BPP} & \textbf{FID} & \textbf{LPIPS} & \textbf{PSNR} \\
\midrule
\multirow[c]{5}{*}{\textbf{BPG}} & 0.056 & 144.322 & 0.460 & 23.771 \\
 & 0.118 & 111.821 & 0.330 & 26.085 \\
 & 0.248 & 81.555 & 0.218 & 28.329 \\
 & 0.503 & 60.512 & 0.127 & 30.864 \\
 & 0.941 & 42.714 & 0.064 & 33.602 \\
\cline{1-5}
\multirow[c]{3}{*}{\textbf{HiFiC}} & 0.220 & 42.610 & 0.055 & 27.713 \\
 & 0.395 & 31.567 & 0.033 & 29.849 \\
 & 0.584 & 25.125 & 0.021 & 31.622 \\
\cline{1-5}
\multirow[c]{5}{*}{\textbf{IPIC}} & 0.095 & 45.367 & 0.211 & 24.470 \\
 & 0.148 & 38.413 & 0.152 & 26.011 \\
 & 0.228 & 36.301 & 0.132 & 27.664 \\
 & 0.550 & 27.075 & 0.074 & 31.178 \\
 & 0.945 & 20.689 & 0.044 & 33.800 \\
\cline{1-5}
\multirow[c]{4}{*}{\textbf{PSC-D}} & 0.019 & 147.317 & 0.591 & 21.648 \\
 & 0.039 & 127.275 & 0.496 & 23.572 \\
 & 0.077 & 105.225 & 0.404 & 25.605 \\
 & 0.155 & 86.787 & 0.316 & 27.711 \\
\cline{1-5}
\multirow[c]{4}{*}{\textbf{PSC-P}} & 0.019 & 67.457 & 0.431 & 18.575 \\
 & 0.039 & 59.119 & 0.353 & 20.178 \\
 & 0.077 & 51.464 & 0.268 & 22.649 \\
 & 0.155 & 44.589 & 0.194 & 25.107 \\
\cline{1-5}
\multirow[c]{13}{*}{\textbf{Ours}} & 0.018 & 68.339 & 0.419 & 19.624 \\
 & 0.023 & 63.731 & 0.365 & 20.566 \\
 & 0.032 & 57.027 & 0.296 & 21.877 \\
 & 0.041 & 52.157 & 0.251 & 22.795 \\
 & 0.053 & 47.834 & 0.220 & 23.401 \\
 & 0.069 & 43.972 & 0.185 & 24.199 \\
 & 0.091 & 40.787 & 0.165 & 24.643 \\
 & 0.122 & 36.890 & 0.134 & 25.587 \\
 & 0.183 & 31.977 & 0.102 & 26.765 \\
 & 0.381 & 28.247 & 0.076 & 27.949 \\
 & 0.594 & 25.705 & 0.063 & 28.694 \\
 & 0.808 & 24.165 & 0.055 & 29.211 \\
 & 1.021 & 22.938 & 0.050 & 29.596 \\
\bottomrule
\end{tabular}
    \label{tab:numerical_compression_imagenet}
\end{table}
\begin{table}[t]
    \centering
\caption{Compression quantitative evaluations, for the Kodak24 $512\times512$ dataset.}
\begin{tabular}{lcccc}
\toprule
\textbf{Method} & \textbf{BPP} & \textbf{FID} & \textbf{LPIPS} & \textbf{PSNR} \\
\midrule
\multirow[c]{7}{*}{\textbf{BPG}} & 0.037 & 148.306 & 0.540 & 24.211 \\
 & 0.044 & 141.575 & 0.513 & 24.638 \\
 & 0.062 & 128.095 & 0.461 & 25.539 \\
 & 0.088 & 114.256 & 0.407 & 26.492 \\
 & 0.121 & 104.289 & 0.357 & 27.375 \\
 & 0.167 & 92.423 & 0.307 & 28.346 \\
 & 0.227 & 82.184 & 0.260 & 29.359 \\
\cline{1-5}
\multirow[c]{2}{*}{\textbf{CRDR-D}} & 0.114 & 94.475 & 0.257 & 27.988 \\
 & 0.210 & 76.015 & 0.167 & 30.069 \\
\cline{1-5}
\multirow[c]{2}{*}{\textbf{CRDR-R}} & 0.114 & 35.676 & 0.094 & 27.315 \\
 & 0.210 & 28.391 & 0.057 & 29.373 \\
\cline{1-5}
\multirow[c]{2}{*}{\textbf{HiFiC}} & 0.191 & 31.713 & 0.068 & 27.376 \\
 & 0.363 & 25.734 & 0.042 & 29.466 \\
\cline{1-5}
\multirow[c]{3}{*}{\textbf{ILLM}} & 0.085 & 36.080 & 0.110 & 25.683 \\
 & 0.159 & 28.556 & 0.072 & 27.296 \\
 & 0.304 & 24.448 & 0.044 & 29.395 \\
\cline{1-5}
\multirow[c]{3}{*}{\textbf{PSC-P}} & 0.025 & 47.334 & 0.473 & 18.516 \\
 & 0.049 & 40.691 & 0.371 & 20.080 \\
 & 0.095 & 35.314 & 0.266 & 21.683 \\
\cline{1-5}
\multirow[c]{2}{*}{\textbf{PerCo (SD)}} & 0.033 & 37.019 & 0.307 & 19.017 \\
 & 0.127 & 26.418 & 0.145 & 22.325 \\
\cline{1-5}
\multirow[c]{6}{*}{\textbf{Ours}} & 0.030 & 32.031 & 0.222 & 22.066 \\
 & 0.038 & 29.117 & 0.190 & 22.551 \\
 & 0.050 & 25.647 & 0.161 & 23.013 \\
 & 0.095 & 24.215 & 0.138 & 23.606 \\
 & 0.149 & 23.199 & 0.124 & 24.069 \\
 & 0.309 & 22.260 & 0.108 & 24.665 \\
\cline{1-5}
\makecell[l]{\textbf{SD 2.1} \\ \textbf{Encoder-Decoder} \\ \textbf{Bound}} & -- & -- & 0.071 & 26.428 \\
\bottomrule
\end{tabular}    
    \label{tab:numerical_compression_kodak}
\end{table}
\begin{table}[t]
    \centering
\caption{Compression quantitative evaluations, for the CLIC2020 $512\times512$ dataset.}
\begin{tabular}{lcccc}
\toprule
\textbf{Method} & \textbf{BPP} & \textbf{FID} & \textbf{LPIPS} & \textbf{PSNR} \\
\midrule
\multirow[c]{7}{*}{\textbf{BPG}} & 0.040 & 112.453 & 0.451 & 24.777 \\
 & 0.046 & 106.936 & 0.428 & 25.213 \\
 & 0.064 & 96.809 & 0.380 & 26.149 \\
 & 0.088 & 86.578 & 0.331 & 27.138 \\
 & 0.119 & 77.203 & 0.285 & 28.072 \\
 & 0.160 & 68.827 & 0.242 & 29.049 \\
 & 0.214 & 60.534 & 0.204 & 30.063 \\
\cline{1-5}
\multirow[c]{3}{*}{\textbf{CRDR-D}} & 0.106 & 52.459 & 0.193 & 28.934 \\
 & 0.188 & 40.228 & 0.124 & 31.062 \\
 & 0.355 & 27.793 & 0.070 & 33.705 \\
\cline{1-5}
\multirow[c]{3}{*}{\textbf{CRDR-R}} & 0.106 & 14.292 & 0.074 & 28.352 \\
 & 0.188 & 10.022 & 0.045 & 30.452 \\
 & 0.355 & 6.225 & 0.025 & 33.028 \\
\cline{1-5}
\multirow[c]{2}{*}{\textbf{HiFiC}} & 0.173 & 14.000 & 0.053 & 28.878 \\
 & 0.322 & 10.157 & 0.032 & 31.040 \\
\cline{1-5}
\multirow[c]{4}{*}{\textbf{ILLM}} & 0.081 & 12.666 & 0.087 & 26.898 \\
 & 0.146 & 9.923 & 0.056 & 28.662 \\
 & 0.271 & 8.246 & 0.034 & 30.840 \\
 & 0.391 & 7.222 & 0.024 & 32.233 \\
\cline{1-5}
\multirow[c]{3}{*}{\textbf{PSC-P}} & 0.025 & 15.007 & 0.435 & 18.153 \\
 & 0.049 & 13.527 & 0.335 & 20.110 \\
 & 0.095 & 11.879 & 0.229 & 22.276 \\
\cline{1-5}
\multirow[c]{2}{*}{\textbf{PerCo (SD)}} & 0.033 & 13.896 & 0.287 & 18.111 \\
 & 0.127 & 7.888 & 0.128 & 22.453 \\
\cline{1-5}
\multirow[c]{6}{*}{\textbf{Ours}} & 0.030 & 9.459 & 0.186 & 22.630 \\
 & 0.038 & 8.386 & 0.160 & 23.171 \\
 & 0.050 & 7.755 & 0.137 & 23.748 \\
 & 0.095 & 7.340 & 0.116 & 24.472 \\
 & 0.149 & 7.164 & 0.103 & 25.008 \\
 & 0.309 & 6.825 & 0.088 & 25.782 \\
\cline{1-5}
\makecell[l]{\textbf{SD 2.1} \\ \textbf{Encoder-Decoder} \\ \textbf{Bound}} & -- & -- & 0.056 & 27.901 \\
\bottomrule
\end{tabular}    
    \label{tab:numerical_compression_clic}
\end{table}
\begin{table}[t]
    \centering
\caption{Compression quantitative evaluations, for the DIV2K $512\times512$ dataset.}
\begin{tabular}{lcccc}
\toprule
\textbf{Method} & \textbf{BPP} & \textbf{FID} & \textbf{LPIPS} & \textbf{PSNR} \\
\midrule
\multirow[c]{5}{*}{\textbf{BPG}} & 0.058 & 152.623 & 0.490 & 22.793 \\
 & 0.068 & 146.219 & 0.463 & 23.217 \\
 & 0.096 & 134.029 & 0.407 & 24.124 \\
 & 0.134 & 122.895 & 0.348 & 25.095 \\
 & 0.327 & 87.879 & 0.192 & 28.049 \\
\cline{1-5}
\multirow[c]{2}{*}{\textbf{CRDR-D}} & 0.152 & 89.766 & 0.203 & 26.618 \\
 & 0.274 & 69.466 & 0.118 & 28.762 \\
\cline{1-5}
\multirow[c]{2}{*}{\textbf{CRDR-R}} & 0.152 & 39.291 & 0.081 & 26.110 \\
 & 0.274 & 31.422 & 0.048 & 28.275 \\
\cline{1-5}
\textbf{HiFiC} & 0.226 & 39.832 & 0.066 & 26.138 \\
\cline{1-5}
\multirow[c]{3}{*}{\textbf{ILLM}} & 0.104 & 42.591 & 0.107 & 24.579 \\
 & 0.187 & 34.103 & 0.069 & 26.206 \\
 & 0.332 & 29.267 & 0.043 & 28.229 \\
\cline{1-5}
\multirow[c]{3}{*}{\textbf{PSC-P}} & 0.025 & 48.350 & 0.474 & 15.869 \\
 & 0.049 & 44.267 & 0.386 & 17.470 \\
 & 0.095 & 38.406 & 0.281 & 19.220 \\
\cline{1-5}
\multirow[c]{2}{*}{\textbf{PerCo (SD)}} & 0.033 & 40.944 & 0.325 & 16.518 \\
 & 0.127 & 29.516 & 0.155 & 20.483 \\
\cline{1-5}
\multirow[c]{6}{*}{\textbf{Ours}} & 0.030 & 36.283 & 0.238 & 19.769 \\
 & 0.038 & 32.917 & 0.206 & 20.319 \\
 & 0.050 & 30.724 & 0.177 & 20.881 \\
 & 0.095 & 29.322 & 0.150 & 21.566 \\
 & 0.149 & 28.011 & 0.132 & 22.107 \\
 & 0.309 & 26.756 & 0.114 & 22.867 \\
\cline{1-5}
\makecell[l]{\textbf{SD 2.1} \\ \textbf{Encoder-Decoder} \\ \textbf{Bound}} & -- & -- & 0.076 & 24.878 \\
\bottomrule
\end{tabular}    
    \label{tab:numerical_compression_div2k}
\end{table}
\begin{table}[t]
    \centering
\caption{Compression quantitative evaluations, for the CLIC2020 $768\times768$ dataset.}
    \begin{tabular}{lcccc}
\toprule
\textbf{Method} & \textbf{BPP} & \textbf{FID} & \textbf{LPIPS} & \textbf{PSNR} \\
\midrule
\multirow[t]{7}{*}{\textbf{BPG}} & 0.032 & 115.009 & 0.451 & 25.547 \\
\textbf{} & 0.038 & 109.587 & 0.429 & 25.984 \\
\textbf{} & 0.054 & 99.482 & 0.385 & 26.925 \\
\textbf{} & 0.075 & 89.013 & 0.338 & 27.915 \\
\textbf{} & 0.102 & 79.456 & 0.296 & 28.840 \\
\textbf{} & 0.138 & 71.107 & 0.255 & 29.811 \\
\textbf{} & 0.185 & 62.617 & 0.219 & 30.812 \\
\cline{1-5}
\textbf{HiFiC} & 0.154 & 9.359 & 0.052 & 29.745 \\
\cline{1-5}
\multirow[t]{4}{*}{\textbf{ILLM}} & 0.006 & 62.327 & 0.447 & 21.304 \\
\textbf{} & 0.009 & 41.479 & 0.340 & 22.703 \\
\textbf{} & 0.072 & 7.852 & 0.084 & 27.808 \\
\textbf{} & 0.134 & 5.966 & 0.055 & 29.567 \\
\cline{1-5}
\multirow[t]{3}{*}{\textbf{PerCo (SD)}} & 0.003 & 30.409 & 0.517 & 15.339 \\
\textbf{} & 0.032 & 12.869 & 0.269 & 19.018 \\
\textbf{} & 0.126 & 5.419 & 0.122 & 23.387 \\
\cline{1-5}
\multirow[t]{9}{*}{\textbf{Ours}} & 0.007 & 23.862 & 0.404 & 19.672 \\
\textbf{} & 0.008 & 19.521 & 0.354 & 20.532 \\
\textbf{} & 0.010 & 15.559 & 0.314 & 21.207 \\
\textbf{} & 0.014 & 11.362 & 0.262 & 22.116 \\
\textbf{} & 0.017 & 8.722 & 0.227 & 22.722 \\
\textbf{} & 0.022 & 6.753 & 0.192 & 23.366 \\
\textbf{} & 0.042 & 5.051 & 0.156 & 24.136 \\
\textbf{} & 0.066 & 4.549 & 0.133 & 24.739 \\
\textbf{} & 0.137 & 4.132 & 0.108 & 25.650 \\
\cline{1-5}
\makecell[l]{\textbf{SD 2.1} \\ \textbf{Encoder-Decoder} \\ \textbf{Bound}} & -- & -- & 0.055 & 28.850 \\
\bottomrule
\end{tabular}
    \label{tab:numerical_compression_clic_768}
\end{table}
\begin{table}[t]
    \centering
\caption{Compression quantitative evaluations, for the DIV2K $768\times768$ dataset.}
    \begin{tabular}{lcccc}
        \toprule
        \textbf{Method} & \textbf{BPP} & \textbf{FID} & \textbf{LPIPS} & \textbf{PSNR} \\
        \midrule
        \multirow[t]{4}{*}{\textbf{BPG}} & 0.049 & 139.204 & 0.485 & 23.504 \\
        \textbf{} & 0.058 & 132.393 & 0.460 & 23.928 \\
        \textbf{} & 0.081 & 120.682 & 0.406 & 24.841 \\
        \textbf{} & 0.114 & 107.472 & 0.350 & 25.826 \\
        \cline{1-5}
        \textbf{HiFiC} & {0.205} & 23.077 & 0.063 & 27.096 \\
        \cline{1-5}
        \multirow[t]{4}{*}{\textbf{ILLM}} & {0.007} & 92.242 & 0.485 & 19.234 \\
        \textbf{} & {0.011} & 71.777 & 0.369 & 20.522 \\
        \textbf{} & {0.093} & 24.253 & 0.102 & 25.446 \\
        \textbf{} & {0.169} & 18.386 & 0.067 & 27.085 \\
        \cline{1-5}
        \multirow[t]{3}{*}{\textbf{PerCo (SD)}} & {0.003} & 51.915 & 0.519 & 14.436 \\
        \textbf{} & {0.032} & 31.378 & 0.311 & 17.364 \\
        \textbf{} & {0.126} & 16.487 & 0.148 & 21.317 \\
        \cline{1-5}
        \multirow[t]{9}{*}{\textbf{Ours}} & {0.007} & 48.315 & 0.469 & 17.194 \\
        \textbf{} & {0.008} & 42.932 & 0.422 & 17.858 \\
        \textbf{} & {0.010} & 37.680 & 0.383 & 18.413 \\
        \textbf{} & {0.014} & 31.038 & 0.328 & 19.246 \\
        \textbf{} & {0.017} & 26.599 & 0.286 & 19.832 \\
        \textbf{} & {0.022} & 22.073 & 0.247 & 20.432 \\
        \textbf{} & {0.042} & 18.330 & 0.201 & 21.167 \\
        \textbf{} & {0.066} & 16.396 & 0.172 & 21.780 \\
        \textbf{} & {0.137} & 14.662 & 0.138 & 22.707 \\
        \cline{1-5}
        \makecell[l]{\textbf{SD 2.1} \\ \textbf{Encoder-Decoder} \\ \textbf{Bound}} & -- & -- & 0.071 & 25.919 \\
        \bottomrule
    \end{tabular}
    \label{tab:numerical_compression_div2k_768}
\end{table}
\clearpage
\section{Compressed Conditional Generation Supplementary}\label{appendix:cond_compression}
\subsection{Background and Proof of~\Cref{prob:ode_convergence}}
We will prove that~\Cref{prob:ode_convergence} holds for any score-based diffusion model~\citep{song2020score}.
For completeness, we first provide the necessary mathematical background and then proceed to the proof of the proposition.
\subsubsection{Background}

\paragraph{Score-Based Generative Models.}Score-based generative models~\citep{song2020score} define a diffusion process $\smash{\{\rvx(t):t\in [0,T]\}}$, where $p_{0}$ and $p_{T}$ denote the data distribution and the prior distribution, respectively, and $p_{t}$ denotes the distribution of $x(t)$.
Such a diffusion process can generally be modeled as the stochastic differential equation (SDE)
\begin{align}
\delt\rvx=f(\rvx,t)\delt t+g(t)\delt\rvw,\label{eq:forward_sde}
\end{align}
where $f(\cdot,t)$ is called the \emph{drift} coefficient, $g(t)$ is called the \emph{diffusion} coefficient, $\rvw(t)$ is a standard Wiener process, and $\delt t$ denotes an infinitesimal timestep.
Samples from the data distribution $p_{0}$ can be generated by solving the reverse-time SDE~\citep{ANDERSON1982313},
\begin{align}
    &\delt\rvx=\left[f(\rvx,t)-g^{2}(t)\vs_{t}(\rvx)\right]\delt t+g(t)\delt\bar{\rvw},\label{eq:reverse_sde}
\end{align}
starting from samples of $\rvx(T)$.
Here, $\vs_{t}(\rvx(t))\coloneqq\nabla_{\rvx(t)}\log{p_{t}(\rvx(t))}$ is the \emph{score} of $p_{t}$, $\bar{\rvw}(t)$ denotes a standard Wiener process where time flows backwards, and $\delt t$ is an infinitesimal \emph{negative} timestep.
Samples from the data distribution can also be generated by solving the \emph{probability flow} ODE,
\begin{align}
    &\delt\rvx=\left[f(\rvx,t)-\frac{1}{2}g^{2}(t)\vs_{t}(\rvx)\right]\delt t.\label{eq:reverse_ode}
\end{align}

\paragraph{Solving The Reverse-Time SDE.}The reverse-time SDE in~\Cref{eq:reverse_sde} can be solved with any numerical SDE solver (e.g., Euler-Maruyama), which corresponds to some time discretization of the forward and reverse stochastic dynamics.
For the sake of our proof, we adopt the simple solver proposed by~\citet{song2020score},
\begin{align}
    &\rvx_{i-1}=\rvx_{i}-f_{i}(\rvx_{i})+g_{i}^{2}\rvs_{i}(\rvx_{i})+g_{i}\rvz_{i},\quad\rvz_{i}\sim\mathcal{N}(\vzero,\mI),\label{eq:song_discretization}
\end{align}
where $i=T,\hdots,1$ and $f_{i}$ and $g_{i}$ are the time-discretized versions of $f$ and $g$, respectively.
Note that DDPMs~\citep{ho2020denoising} are score-based diffusion models that solve a reverse-time Variance Preserving (VP) SDE, where $f(\rvx(t),t)=-\frac{1}{2}\beta(t)\rvx(t)$ and $g(t)=\sqrt{\beta(t)}$ for some function $\beta$.

\subsubsection{Proof of~\Cref{prob:ode_convergence}}
Given any general score-based diffusion model, we can write the DDCM compressed conditional generation process as
\begin{align}
    &\rvx_{i-1}=\rvx_{i}-f_{i}(\rvx_{i})+g_{i}^{2}\rvs_{i}(\rvx_{i})+g_{i}\gC_{i}(k_{i}),\label{eq:score-based-ddcm}
\end{align}
where $k_{i}$ are picked according to~\Cref{eq:k_choose_conditional}.
Choosing $\gL=\gL_{\text{P}}$, we have
\begin{align}
    k_{i}=\argmin_{k\in\{1,\hdots,K\}}\norm{\gC_{i}(k)-g_{i}\nabla_{\rvx_{i}}\log{p_{i}(\rvy|\rvx_{i})}}^{2}.
\end{align}
Since each $\gC_{i}$ contains $K$ independent samples drawn from a normal distribution $\gN(\vzero,\mI)$, we have
\begin{align}
    \{\gC_{i}(1),\hdots,\gC_{i}(K)\}\underset{K\rightarrow\infty}{\longrightarrow}\mathbb{R}^{n},
\end{align}
where $n$ denotes the dimensionality of each vector in $\gC_{i}$, and $\{\gC_{i}(1),\hdots,\gC_{i}(K)\}$ is the set comprised of all the elements in the $\gC_{i}$ (without repetition).
Since $g_{i}\nabla_{\rvx_{i}}\log{p_{i}(\rvy|\rvx_{i})}\in\mathbb{R}^{n}$, we have
\begin{align}
    \min_{k\in\{1,\hdots,K\}}\norm{\gC_{i}(k)-g_{i}\nabla_{\rvx_{i}}\log{p_{i}(\rvy|\rvx_{i})}}^{2}\underset{K\rightarrow\infty}{\longrightarrow}0.
\end{align}
Thus,
\begin{align}
    \gC_{i}(k_{i})\underset{K\rightarrow\infty}{\longrightarrow}g_{i}\nabla_{\rvx_{i}}\log{p_{i}(\rvy|\rvx_{i})}.\label{eq:noise_convergence_to_grad}
\end{align}
Plugging~\Cref{eq:noise_convergence_to_grad} into~\Cref{eq:score-based-ddcm}, we get
\begin{align}
    \rvx_{i-1}\underset{K\rightarrow\infty}{\longrightarrow}&\rvx_{i}-f_{i}(\rvx_{i})+g_{i}^{2}\rvs_{i}(\rvx_{i})+g_{i}^{2}\nabla_{\rvx_{i}}\log{p_{i}(\rvy|\rvx_{i})}\\
    =&\rvx_{i}-f_{i}(\rvx_{i})+g_{i}^{2}\nabla_{\rvx_{i}}\log{p_{i}(\rvx_{i})}+g_{i}^{2}\nabla_{\rvx_{i}}\log{p_{i}(\rvy|\rvx_{i})}\\
    =&\rvx_{i}-f_{i}(\rvx_{i})+g_{i}^{2}\left[\nabla_{\rvx_{i}}\log{p_{i}(\rvx_{i})}+\nabla_{\rvx_{i}}\log{p_{i}(\rvy|\rvx_{i})}\right]\\
    =&\rvx_{i}-f_{i}(\rvx_{i})+g_{i}^{2}\nabla_{\rvx_{i}}\log{p_{i}(\rvx_{i}|\rvy)},\label{eq:bayes_rule_for_ode}
\end{align}
where in~\Cref{eq:bayes_rule_for_ode} we used Bayes rule and the fact that $\nabla_{\rvx_{i}}\log{p(\rvy)}=0$.
Note that~\Cref{eq:bayes_rule_for_ode} resembles a time discretization of a probability flow ODE (\Cref{eq:reverse_ode}) over the posterior distribution $p_{0}(\rvx_{0}|\rvy)$, with $f(\cdot,t)$ and $\sqrt{2}g(t)$ being the drift and diffusion coefficients in continuous time, respectively.
Thus, when $K\rightarrow\infty$, our compressed conditional generation process becomes a sampler from $p_{0}(\rvx_{0}|\rvy)$.


\subsection{Image Compression as a Private Case of Compressed Conditional Generation}\label{appendix:compression_private_case}
We show that our standard image compression scheme from~\Cref{section:compression} is a private case of our compressed conditional generation scheme from~\Cref{sec:compressed_conditional_generation}, where $\rvy=\rvx_{0}$ and $\gL=\gL_{\text{P}}$.
When $\rvy=\rvx_{0}$ we have
\begin{align}
    \nabla_{\rvx_{i}}\log{p_{i}(\rvy|\rvx_{i})}&=\nabla_{\rvx_{i}}\log{p_{i}(\rvx_{0}|\rvx_{i})}\nonumber\\
    &=\nabla_{\rvx_{i}}\log{p_{i}(\rvx_{i}|\rvx_{0})}-\nabla_{\rvx_{i}}\log{p_{i}(\rvx_{i})}\\
    &=\nabla_{\rvx_{i}}\log{p_{i}(\rvx_{i}|\rvx_{0})}-\vs_{i}(\rvx_{i})\label{eq:compression-grad}
\end{align}
where $\nabla_{\rvx_{i}}\log{p_{i}(\rvx_{i}|\rvx_{0})}$ can be computed in closed-form via the forward diffusion process in~\Cref{eq:forward_gaussian_diffusion}.
In particular, we have $p_{i}(\rvx_{i}|\rvx_{0})=\gN(\rvx_{i};\sqrt{\bar{\alpha}_{i}}\rvx_{0},(1-\bar{\alpha}_{i})\mI)$~\citep{ho2020denoising}, and thus
\begin{align}
    \nabla_{\rvx_{i}}\log{p_{i}(\rvx_{i}|\rvx_{0})}&=-\nabla_{\rvx_{i}}\frac{\norm{\rvx_{i}-\sqrt{\bar{\alpha}_{i}}\rvx_{0}}^{2}}{2(1-\bar{\alpha}_{i})}\\
    &=-\frac{\rvx_{i}-\sqrt{\bar{\alpha}_{i}}\rvx_{0}}{1-\bar{\alpha}_{i}}\\
    &=\frac{\sqrt{\bar{\alpha}_{i}}\rvx_{0}-\rvx_{i}}{1-\bar{\alpha}_{i}}.\label{eq:vp-sde-forward-grad}
\end{align}
Moreover, it is well known that~\citep{ho2020denoising,song2020score}
\begin{align}
    \vs_{i}(\rvx_{i})=\frac{\sqrt{\bar{\alpha}_{i}}\hat{\rvx}_{0|i}-\rvx_{i}}{1-\bar{\alpha}_{i}},\label{eq:score-mmse-in-vp-sde}
\end{align}
where $\hat{\rvx}_{0|i}$ is the (time-aware) Minimum Mean-Squared-Error (MMSE) estimator of $\rvx_{0}$ given $\rvx_{i}$.
Plugging~\Cref{eq:score-mmse-in-vp-sde,eq:vp-sde-forward-grad} into~\cref{eq:compression-grad}, we get
\begin{align}
    \nabla_{\rvx_{i}}\log{p_{i}(\rvx_{0}|\rvx_{i})}=\frac{\sqrt{\bar{\alpha}_{i}}}{1-\bar{\alpha}_{i}}(\rvx_{0}-\hat{\rvx}_{0|i}).\label{eq:compression_vp_sde_likelihood}
\end{align}
Thus, we have
\begin{align}
    k_{i}&=\argmin_{k\in\{1,\hdots,K\}}\gL_{\text{P}}(\rvy,\rvx_{i},\gC_{i},k)\label{eq:compression_posterior}\\
    &=\argmin_{k\in\{1,\hdots,K\}}\bignorm{\gC_{i}(k)-\sigma_{i}\frac{\sqrt{\bar{\alpha}_{i}}}{1-\bar{\alpha}_{i}}(\rvx_{0}-\hat{\rvx}_{0|i})}^{2}\\
    &=\argmin_{k\in\{1,\hdots,K\}}\bignorm{\gC_{i}(k)}^{2}-2\langle\gC_{i}(k),\sigma_{i}\frac{\sqrt{\bar{\alpha}_{i}}}{1-\bar{\alpha}_{i}}(\rvx_{0}-\hat{\rvx}_{0|i})\rangle+\bignorm{\sigma_{i}\frac{\sqrt{\bar{\alpha}_{i}}}{1-\bar{\alpha}_{i}}(\rvx_{0}-\hat{\rvx}_{0|i})}^{2}\\
    &=\argmin_{k\in\{1,\hdots,K\}}\bignorm{\gC_{i}(k)}^{2}-2\langle\gC_{i}(k),\sigma_{i}\frac{\sqrt{\bar{\alpha}_{i}}}{1-\bar{\alpha}_{i}}(\rvx_{0}-\hat{\rvx}_{0|i})\rangle.
\end{align}
Below, we show that $\norm{\gC_{i}(k)}^{2}$ is roughly equal for every $k$.
Thus, it holds that
\begin{align}
    k_{i}&=\argmin_{k\in\{1,\hdots,K\}}\bignorm{\gC_{i}(k)}^{2}-2\langle\gC_{i}(k),\sigma_{i}\frac{\sqrt{\bar{\alpha}_{i}}}{1-\bar{\alpha}_{i}}(\rvx_{0}-\hat{\rvx}_{0|i})\rangle\\
    &\approx\argmin_{k\in\{1,\hdots,K\}}\text{const}-2\langle\gC_{i}(k),\sigma_{i}\frac{\sqrt{\bar{\alpha}_{i}}}{1-\bar{\alpha}_{i}}(\rvx_{0}-\hat{\rvx}_{0|i})\rangle\\
    &=\argmax_{k\in\{1,\hdots,K\}}\langle\gC_{i}(k),\sigma_{i}\frac{\sqrt{\bar{\alpha}_{i}}}{1-\bar{\alpha}_{i}}(\rvx_{0}-\hat{\rvx}_{0|i})\rangle\\
    &=\argmax_{k\in\{1,\hdots,K\}}\sigma_{i}\frac{\sqrt{\bar{\alpha}_{i}}}{1-\bar{\alpha}_{i}}\langle\gC_{i}(k),\rvx_{0}-\hat{\rvx}_{0|i}\rangle\\
    &=\argmax_{k\in\{1,\hdots,K\}}\langle\gC_{i}(k),\rvx_{0}-\hat{\rvx}_{0|i}\rangle.\label{eq:equiv_kstar_compression}
\end{align}
Note that the noise selection strategy in~\Cref{eq:equiv_kstar_compression} is similar to that of our standard compression scheme, namely~\Cref{eq:compression_rule}.
Thus, our compression method is a private case of our compressed conditional generation approach.
In practice, we used~\Cref{eq:compression_rule} instead of~\Cref{eq:compression_posterior} since the former worked slightly better.

To show that $\norm{\gC_{i}(k)}^{2}$ is roughly constant for every $k$, note that $\gC_{i}(k)$ is a sample from a $n$-dimensional multivariate normal distribution $\mathcal{N}(\vzero,\mI)$.
Thus, $\norm{\gC_{i}(k)}^{2}$ is a sample from a chi-squared distribution with $n$ degrees of freedom.
It is well known that samples from this distribution strongly concentrate around its mean $n$ for large values of $n$.
Namely, $\norm{\gC_{i}(k)}^{2}$ is highly likely to be close to $n$, especially for relatively small values of $K$.
\clearpage
\subsection{Compressed Posterior Sampling for Image Restoration}\label{appendix:zero-shot}
\begin{figure*}[t]
    \centering
    \includegraphics[width=1\textwidth]{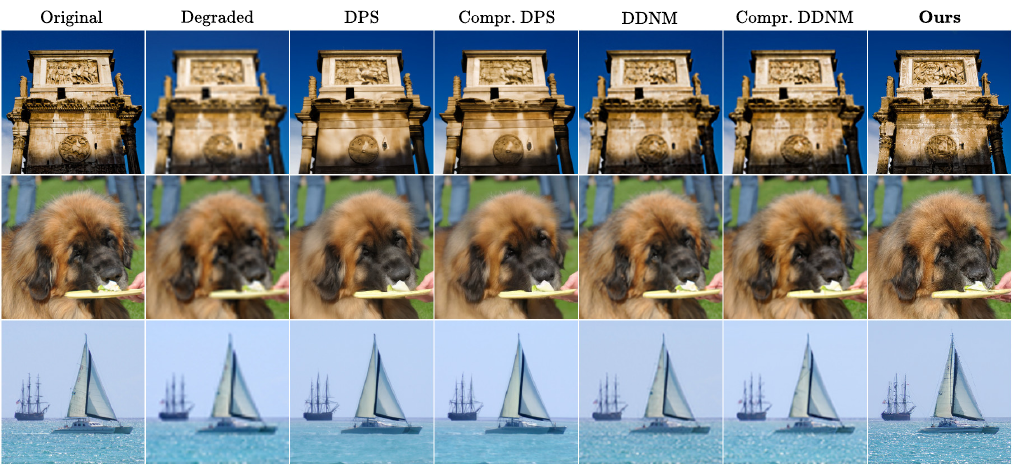}
    \caption{\textbf{Qualitative comparison of zero-shot image super-resolution methods (posterior sampling).} Our approach clearly produces better output perceptual quality compared to previous methods.}
    \label{fig:srx4_additional_samples}
\end{figure*}

\begin{figure*}[t]
    \centering
    \includegraphics[width=1\textwidth]{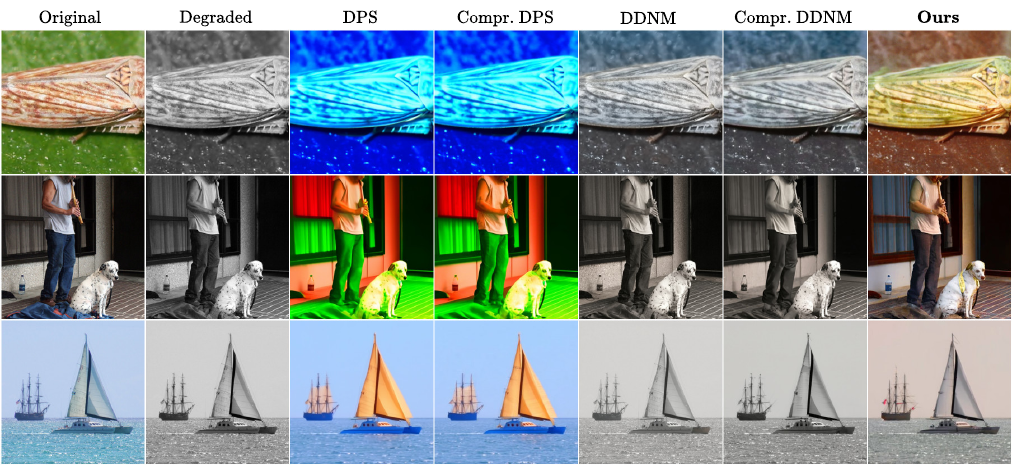}
    \caption{\textbf{Qualitative comparison of zero-shot image colorization methods (posterior sampling).} Our approach clearly produces better output perceptual quality compared to previous methods.}
    \label{fig:colorization_additional_samples}
\end{figure*}

DPS and DDNM are implemented with the official settings recommended by the authors~\citep{chung2023diffusion,wang2023zeroshot}.
Specifically, DPS uses DDPM with $T=1000$ sampling steps, and DDNM uses DDIM with $\eta=0.85$ and $T=100$ sampling steps.
We also tried $T=1000$ for DDNM and found that $T=100$ works slightly better for the tasks considered.
The additional qualitative comparisons in \Cref{fig:srx4_additional_samples,fig:colorization_additional_samples} further demonstrate that our method produces better output perceptual quality compared to DPS and DDNM.
\clearpage
\subsection{Compressed Real-World Face Image Restoration}\label{app:dmax-ot}
\subsubsection{Explaining the Choice of $\vr(\rvy)$}
To explain our choice of $\vr(\rvy)\approx\mathbb{E}[\rvx_{0}|\rvy]$, first note that the MSE of \emph{any} estimator $\hat{\rvx}_{0}$ of $\rvx_{0}$ given $\rvy$ can be written as~\citep{freirich2021a}
\begin{align}
    \text{MSE}(\rvx_{0},\hat{\rvx}_{0})
    &=\text{MSE}(\rvx_{0},\vr(\rvy))+\text{MSE}(\vr(\rvy),\hat{\rvx}_{0})\nonumber\\
    &=\text{MSE}(\vr(\rvy),\hat{\rvx}_{0})+c\label{eq:orthogonality}.
\end{align}
where $c$, the MMSE, does not depend on $\hat{\rvx}_{0}$.
In theory, our goal is to optimize the tradeoff between the MSE of $\hat{\rvx}_0$ and its output perceptual quality according to some quality measure $Q$. This can be accomplished by solving
\begin{align}\label{eq:desired_tradeoff}
    \min_{\hat{\rvx}_{0}}\text{MSE}(\rvx_{0},\hat{\rvx}_{0})+\lambda Q(\hat{\rvx}_{0}),
\end{align}
where $\lambda$ is some hyper-parameter that controls the perception-distortion tradeoff.
At test time, however, we do not have access to the original image $\rvx_0$.
By plugging \Cref{eq:orthogonality} into \Cref{eq:desired_tradeoff} we obtain
\begin{align}
    \min_{\hat{\rvx}_{0}}\text{MSE}(\rvx_{0},\hat{\rvx}_{0})+\lambda Q(\hat{\rvx}_{0})&=\min_{\hat{\rvx}_{0}}\text{MSE}(\vr(\rvy),\hat{\rvx}_{0})+c+\lambda Q(\hat{\rvx}_{0})\nonumber\\
    &=\min_{\hat{\rvx}_{0}}\text{MSE}(\vr(\rvy),\hat{\rvx}_{0})+\lambda Q(\hat{\rvx}_{0}).\label{eq:opt-mse-real-world}
\end{align}
Namely, as long as $\vr(\rvy)$ is a good approximation of the \emph{true} MMSE estimator, we can rely on it for optimizing tradeoff (\ref{eq:desired_tradeoff}) without having access to $\rvx_{0}$.
This resembles our approach in \Cref{sec:bfr}, where we \emph{greedily} optimize \Cref{eq:opt-mse-real-world} throughout the trajectory of the DDCM sampling process.

\subsubsection{Additional Details and Experiments}
\begin{figure}
    \centering
    \includegraphics[height=21cm]{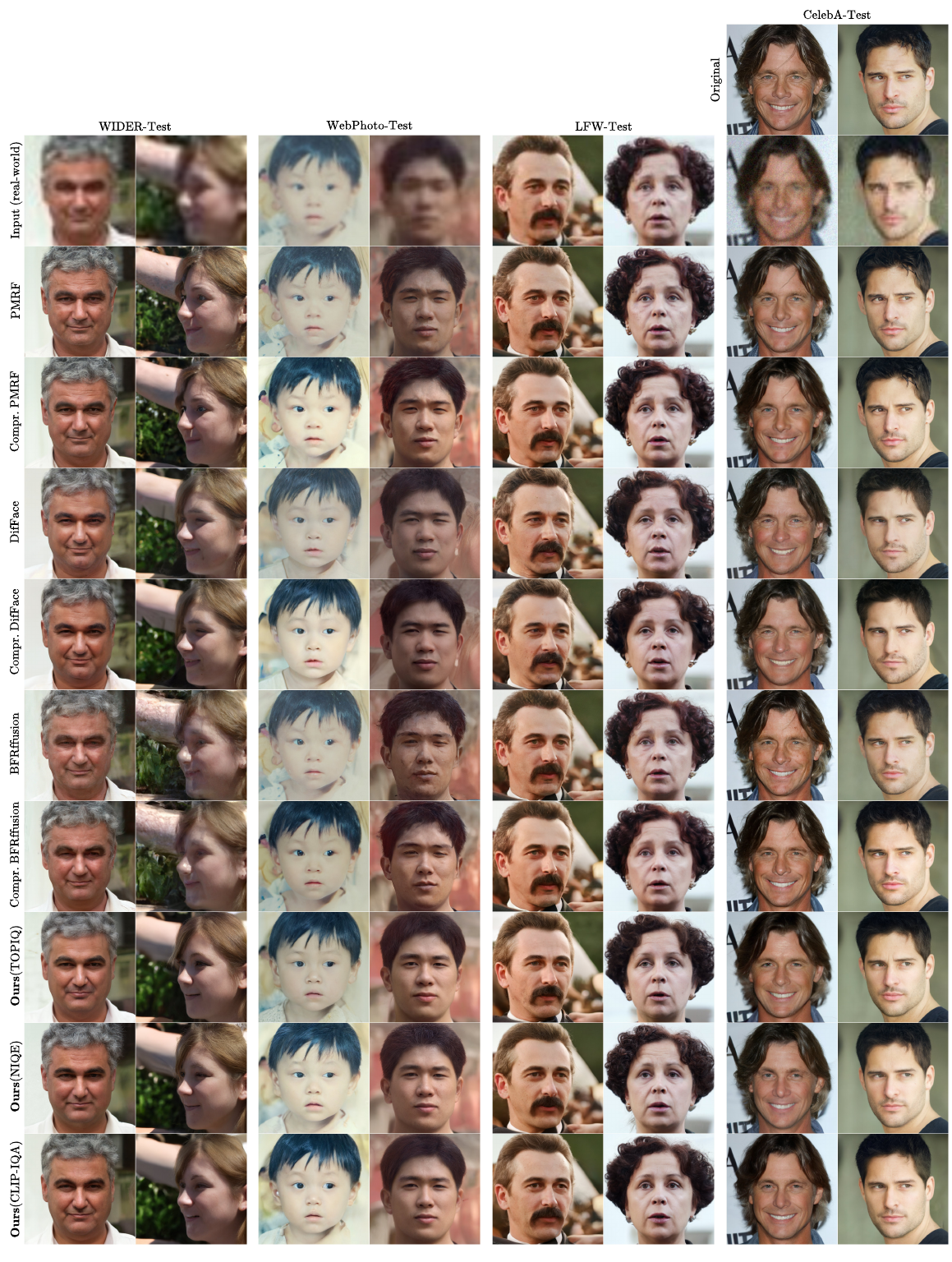}
    \caption{\textbf{Qualitative comparison of real-world face image restoration methods}. Our method produces high perceptual quality results with less artifacts compared to previous methods, especially for challenging datasets such as WIDER-Test.}
    \label{fig:real-world-qualitative-appendix}
\end{figure}

\begin{figure}
    \centering
    \includegraphics[width=1\textwidth]{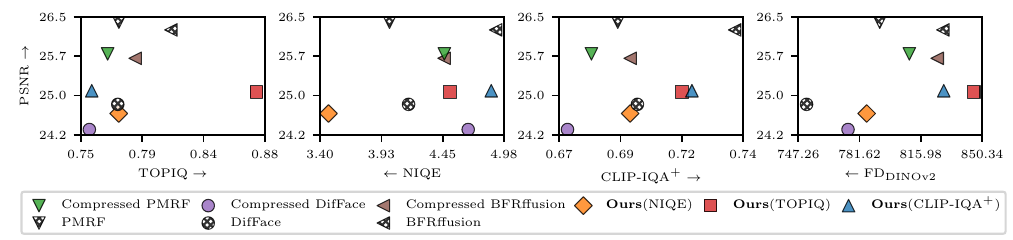}
    \caption{\textbf{Quantitative comparison of real-world face image restoration methods, evaluated on the CelebA-Test dataset.} We successfully optimize each NR-IQA measure, surpassing the scores of previous methods. Here, only our NIQE-based solution generalizes well to $\text{FD}_{\text{DINOv2}}$ in terms of perceptual quality.}
    \label{fig:celeba-quantitative-appendix}
\end{figure}

\begin{figure}
    \centering
    \includegraphics[width=1\textwidth]{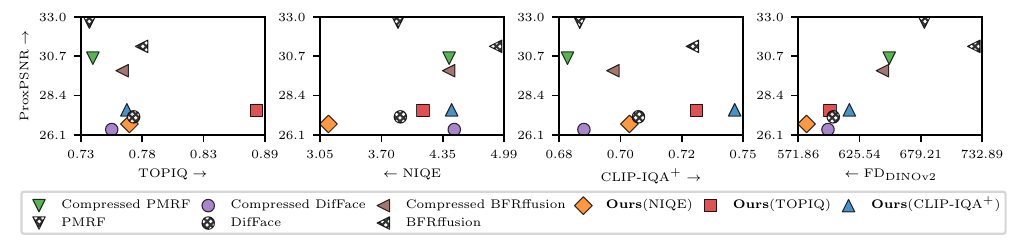}
    \caption{\textbf{Quantitative comparison of real-world face image restoration methods, evaluated on the LFW-Test dataset.} We successfully optimize each NR-IQA measure, surpassing the scores of previous methods. All our solutions achieve impressive $\text{FD}_{\text{DINOv2}}$ scores, while our NIQE-based solution surpasses all methods according to this measure.}
    \label{fig:lfw-quantitative-appendix}
\end{figure}

\begin{figure}
    \centering
    \includegraphics[width=1\textwidth]{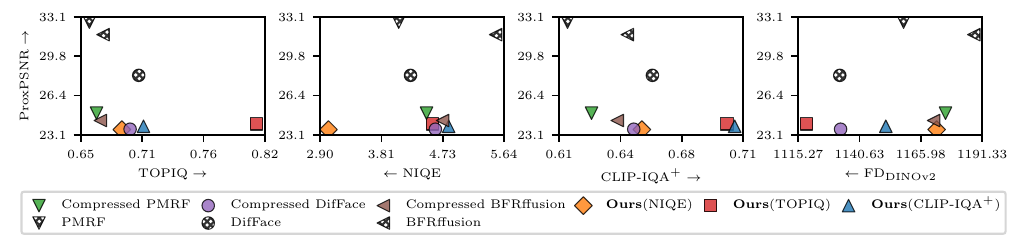}
    \caption{\textbf{Quantitative comparison of real-world face image restoration methods, evaluated on the WebPhoto-Test dataset.} We successfully optimize each NR-IQA measure, surpassing the scores of previous methods. Our TOPIQ-based solution achieves the best $\text{FD}_{\text{DINOv2}}$ scores compared to all methods..}
    \label{fig:webphoto-quantitative-appendix}
\end{figure}

We use the PyIQA package to compute all perceptual quality measures, with \texttt{niqe} for NIQE, \texttt{topiq\_nr-face} for TOPIQ, \texttt{clipiqa+} for $\text{CLIP-IQA}^{+}$, and \texttt{fid\_dinov2} for $\text{FD}_{\text{DINOv2}}$, adopting the default settings for each measure.
Additional qualitative and quantitative comparisons are shown in \cref{fig:real-world-qualitative-appendix,,fig:lfw-quantitative-appendix,,fig:celeba-quantitative-appendix,,fig:webphoto-quantitative-appendix}.


\clearpage

\subsection{Compressed Classifier Guidance}\label{appendix:classifier-guidance}
\begin{figure*}[t]
    \centering
    \includegraphics[height=15cm]{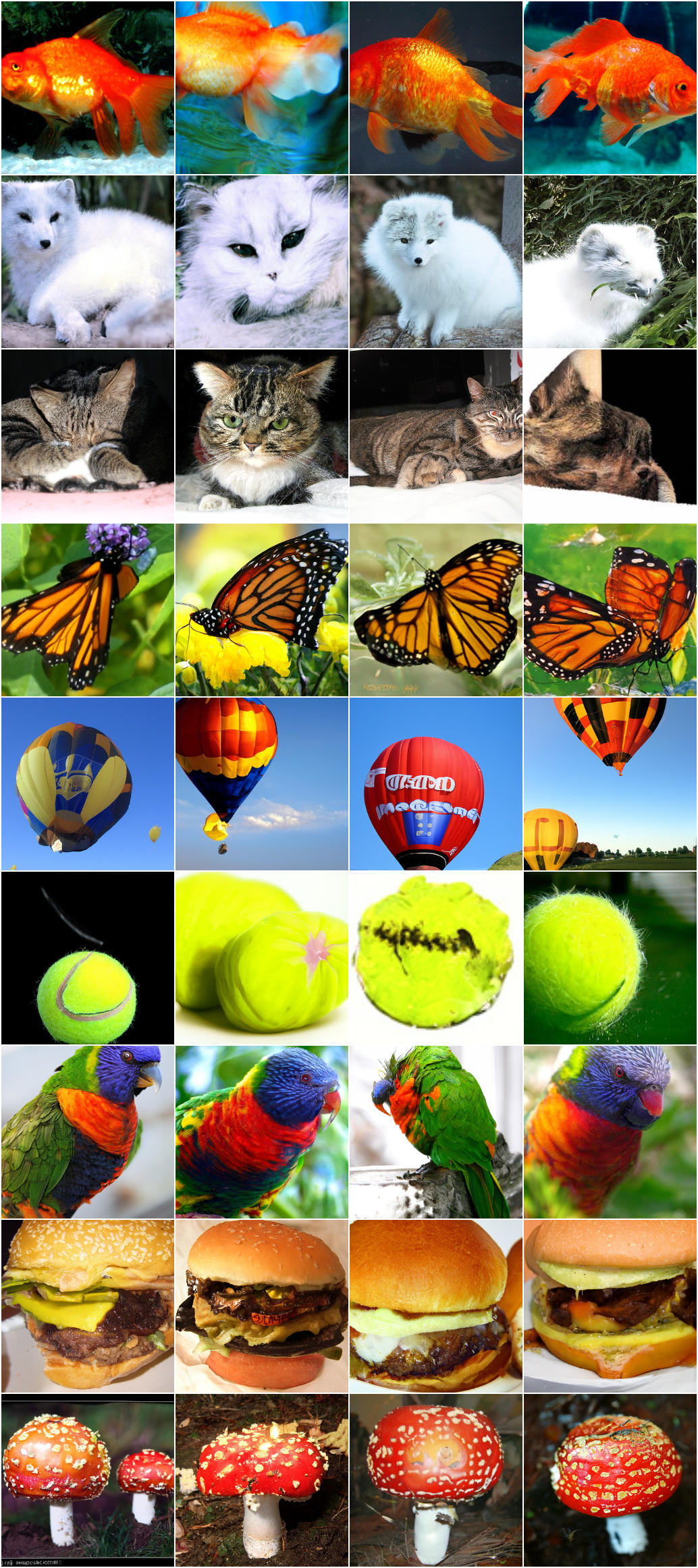}\hspace{1cm}\includegraphics[height=15cm]{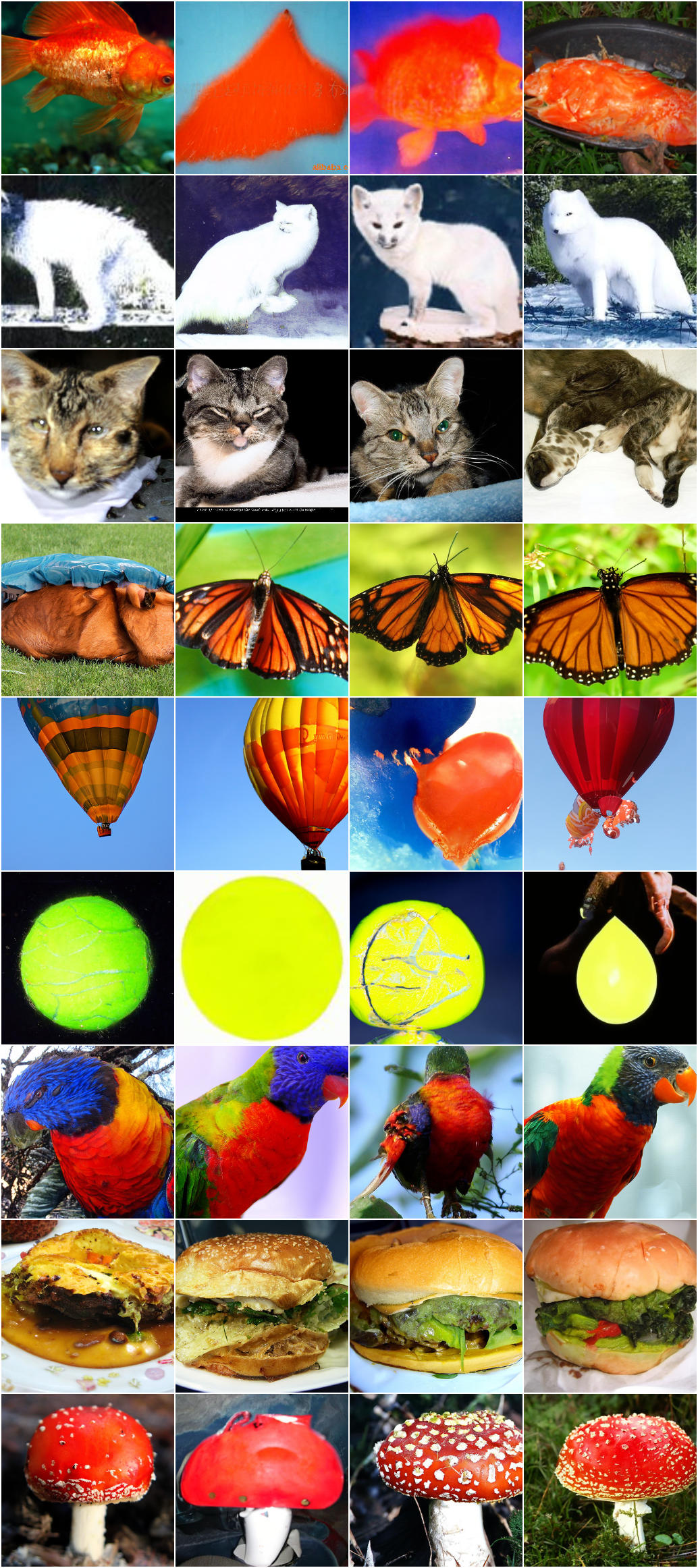}
    \caption{\textbf{Qualitative comparison of CCG (left) with CG (right).} CCG achieves superior image quality compared to CG while avoiding the use of classifier gradients. Additionally, CCG enables decompression without requiring access to the original class labels.}
    \label{fig:cg_visual_comparison}
\end{figure*}
Consider the case where $\rvy$ represents the \emph{class} of an image $\rvx_{0}$.
An unconditional score-based generative model can be \emph{guided} to generate samples the posterior $p_{0}(\rvx_{0}|\rvy)$, by perturbing the generated samples according to the gradient of a time-dependent trained classifier $c_{\theta}(\rvy;\rvx_{i},i)\approx p_{i}(\rvy|\rvx_{i})$~\citep{dhariwal2021diffusion}.
This approach is known as classifier guidance (CG).
Such a guidance method can be interpreted as an attempt to \emph{confuse} the classifier by perturbing its input adversarially~\citep{ho2021classifier}.
However, trained classifiers are typically not robust to adversarial perturbations, making their gradients largely unreliable and unaligned with human perception~\citep{advattacks,tsipras2018robustness,ganz-perceptual}.
Thus, the standard CG approach has not seen major success~\citep{ho2021classifier}.

We propose an alternative to this method, circumventing the reliance on the classifier's gradient.
Specifically, we set $\gL$ in \Cref{eq:k_choose_conditional} as
\begin{align}
\gL(\rvy,\rvx_{i},\gC_{i},k)=-\log{c_{\theta}(\rvy;\vmu_{i}(\rvx_{i})+\sigma_{i}\gC_{i}(k),i)}.\label{eq:classifier-guidance-ours}
\end{align}
Thus, $\gL(\rvy,\rvx_{i},\gC_{i},k)$ attains a lower value when $\sigma_{i}\gC_{i}(k)$ points in some direction that maximizes the probability of the class $\rvy$.
Note that since the codebooks remain fixed, choosing $k_{i}$ (out of $1,\hdots,K$) to minimize~\Cref{eq:k_choose_conditional} would always lead to the same generated sample for every $\rvy$.
Thus, we promote sample diversity by first \emph{randomly} selecting a subset of $\tilde{K}<K$ indices $k_{i,1},\hdots,k_{i,\tilde{K}}\sim\text{Unif}(\{1,\hdots,K\})$, and then choosing
\begin{align}
    k_{i}=\argmin_{k\in\{k_{i,1},\hdots,k_{i,\tilde{K}}\}}\gL(\rvy,\rvx_{i},\gC_{i},k).
\end{align}
We coin our method Compressed CG (CCG).

We compare our proposed CCG with the standard CG using the same unconditional diffusion model and time-dependent classifier trained on ImageNet $256\times256$~\citep{deng2009imagenet,dhariwal2021diffusion}.
We compare the methods ``on the same grounds'' by using the same standard DDPM noise schedule and $T=1000$ diffusion steps.
Our method is assessed with $K=256$ and $\tilde{K}=2$, while the standard CG is assessed with CG scales $s\in\{1,10,20\}$.
The quantitative comparison in~\Cref{tab:cg_comparison} shows that CCG achieves better (lower) FID and $\text{FD}_{\text{DINOv2}}$ scores.
A visual comparison is provided in~\Cref{fig:cg_visual_comparison}.
Note that while using DDCM with standard CG does still produces compressed output images, decoding the produced bit-streams requires access to $\rvy$.
Using DDCM with CCG instead sidesteps this limitation, as $\rvy$ is not needed for decompression.

\begin{table}[ht]
\centering
\caption{Quantitative comparison of \emph{compressed} classifier guidance (CCG) and standard classifier guidance (CG) for ImageNet $256\times 256$ conditional image generation, using an unconditional DDM and a classifier for guidance. Our proposed CCG not only outperforms CG in terms of generation performance, but also automatically produces compressed image representations.}
\begin{tabular}{lcc}
\toprule
 &  \begin{tabular}{c}Compressed CG \textbf{(Ours)}\\$K=256,\tilde{K}=2$\end{tabular}& \begin{tabular}{c}Standard CG\\$s=1\mid10\mid20$\end{tabular} \\
\midrule
FID & $\mathbf{13.669}$ & $31.548\mid14.481\mid 14.921$ \\
$\text{FD}_{\text{DINOv2}}$ & $\mathbf{204.693}$ & $459.42 \mid255.41 \mid248.34 $ \\
\bottomrule
\end{tabular}
\label{tab:cg_comparison}
\end{table}

\clearpage
\subsection{Compressed Classifier-Free Guidance}\label{appendix:classifier-free-guidance}
\begin{figure*}[t]
    \centering
    \includegraphics[width=\textwidth]{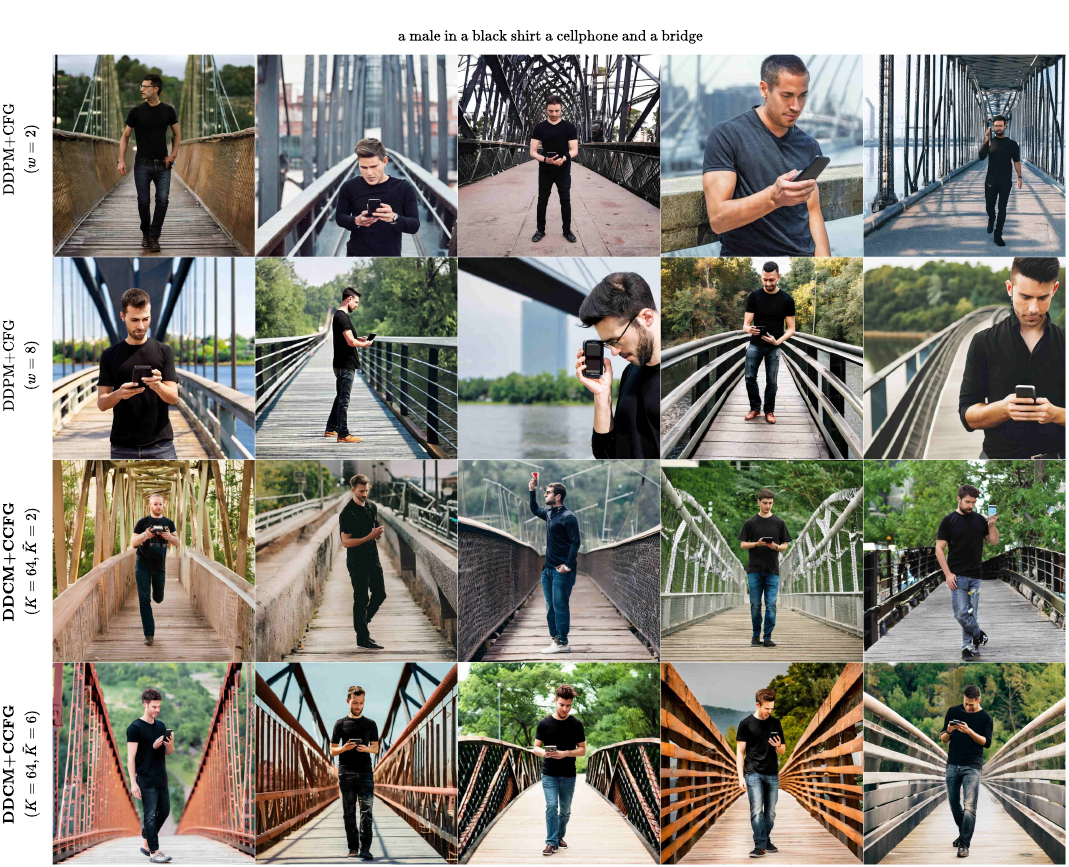}
    \caption{\textbf{Qualitative comparison of CCFG with CFG.} CCFG achieves comparable image quality and diversity to CFG, while enabling decompression without requiring the original inputs.}
    \label{fig:ccfg_visual_1}
\end{figure*}

\begin{figure*}[t]
    \centering
    \includegraphics[width=\textwidth]{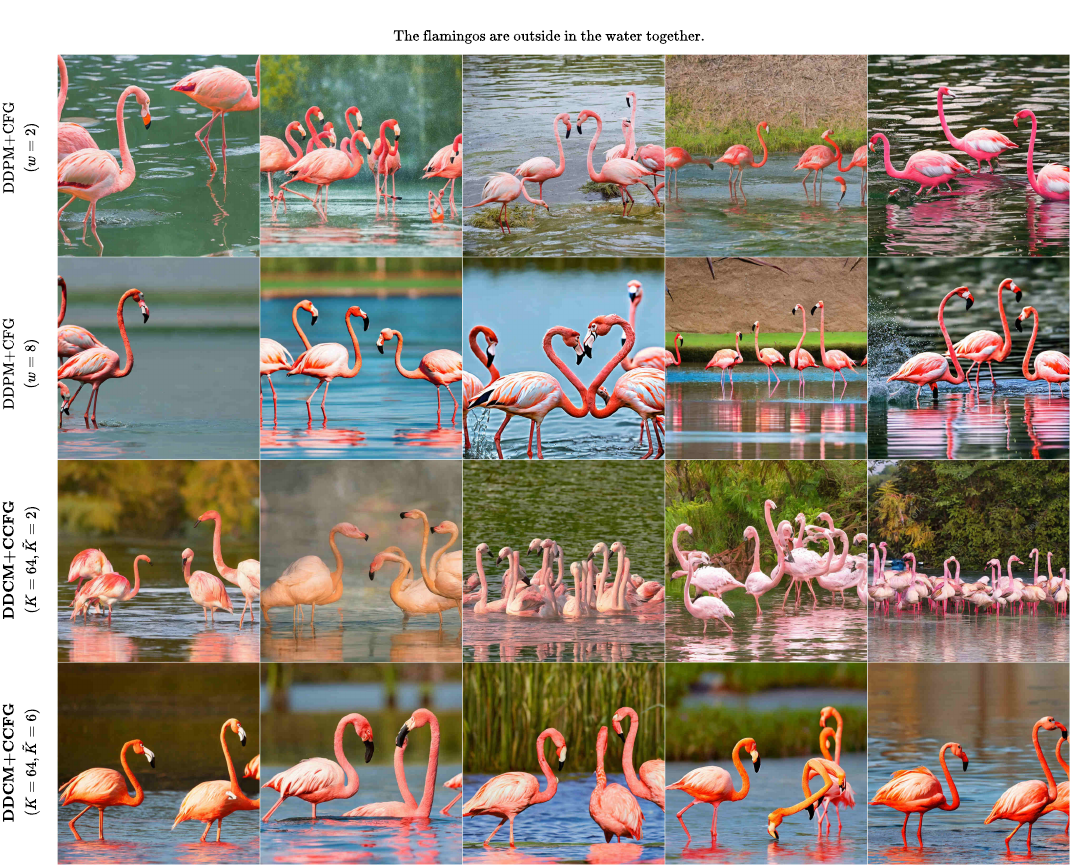}
    \caption{\textbf{Qualitative comparison of CCFG with CFG.} CCFG achieves comparable image quality and diversity to CFG, while enabling decompression without requiring the original inputs.}
    \label{fig:ccfg_visual_2}
\end{figure*}

\begin{figure*}
    \centering
    \includegraphics[width=0.5\textwidth]{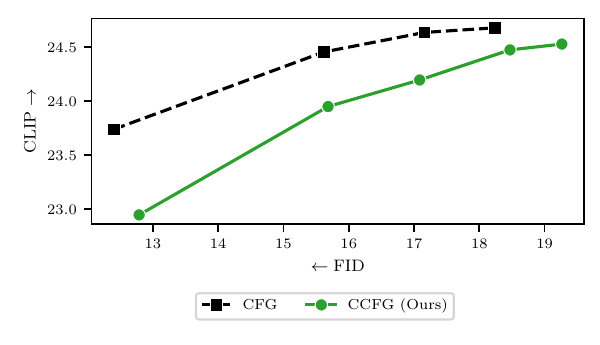}
    \caption{\textbf{Quantitative evaluation of CCFG and CFG.} CCFG achieves comparable FID scores to CFG while achieving slightly lower fidelity to the input prompts. However, unlike CFG, CCFG enables decompression without access to the original conditioning inputs.}
    \label{fig:ccfg-cf-quantitative-comparison}
\end{figure*}
The task of text-conditional image generation can be solved using a \emph{conditional} diffusion model, which, theoretically speaking, learns to sample from the posterior distribution $p_{0}(\rvx_{0}|\rvy)$.
In practice, however, using a conditional model directly typically yields low fidelity to the inputs.
To address this limitation, CG can be used to improve this fidelity at the expense of sample quality and diversity~\citep{dhariwal2021diffusion}.
Classifier-Free Guidance (CFG) is used more often in practice, as it achieves the same tradeoff by mixing the conditional and unconditional scores during sampling~\citep{ho2021classifier}, thus eliminating the need for a classifier.
Particularly, assuming we have access to both the conditional score $\vs_{i}(\rvx_{i},\rvy)\coloneqq\nabla_{\rvx_{i}}\log{p_{i}(\rvx_{i}|\rvy)}$ and the unconditional one $\vs_{i}(\rvx_{i})$, CFG proposes to modify the conditional score by
\begin{align}
    \tilde{\vs}_{i}(\rvx_{i},\rvy)=(1+w)\vs_{i}(\rvx_{i},\rvy)-w\vs_{i}(\rvx_{i}),
\end{align}
where $w$, the CFG scale, is a hyper-parameter controlling the tradeoff between sample quality and diversity.

Here, we introduce a new CFG method that allows generating compressed conditional samples using any pair of conditional and unconditional diffusion models, while controlling the tradeoff between generation quality and the fidelity to the inputs.
Specifically, since $\nabla_{\rvx_{i}}\log{p_{i}(\rvy|\rvx_{i})}=\vs_{i}(\rvx_{i}|\rvy)-\vs_{i}(\rvx_{i})$, we simply use
\begin{align}
    \gL(\rvy,\rvx_{i},\gC_{i},k)=-\langle\gC_{i}(k),\vs_{i}(\rvx_{i}|\rvy)-\vs_{i}(\rvx_{i})\rangle.\label{eq:l_ccfg}
\end{align}
Note that optimizing~\Cref{eq:l_ccfg} is roughly  equivalent to optimizing $\gL_{\text{P}}$ when $\rvx_{i}$ is high dimensional (see~\Cref{appendix:compression_private_case}).
As in~\Cref{appendix:classifier-guidance}, we promote sample diversity by choosing $k_{i}$ from a randomly sampled subset of $\tilde{K}<K$ indices at each step during the generation.
We coin our method Compressed CFG (CCFG).

We implement our method using SD 2.1 trained on $768\times768$ images, adopting a DDPM noise schedule with $T=1000$ diffusion steps, $K=64$ fixed vectors in each codebook and $\tilde{K}\in\{2,3,4,6,9\}$.
We compare against the same diffusion model with standard DDPM sampling, using $T=1000$ steps and CFG scales $w\in\{2,5,8,11\}$.
The generative performance of both methods is assessed by computing the FID between 10k generated samples and MS-COCO, similarly to \Cref{sec:method}.
Additionally, we evaluate the alignment between the outputs and the input text prompts using the CLIP score~\citep{hessel2021clipscore} with the OpenAI CLIP ViT-L/14 model~\citep{pmlr-v139-radford21a}.

Figure~\ref{fig:ccfg-cf-quantitative-comparison} shows that our CCFG method is on par with CFG in terms of FID, while CFG produces higher CLIP scores.
This suggests that the outputs of CFG better align with the input text prompts compared to CCFG.
Yet, the qualitative comparisons in \Cref{fig:ccfg_visual_1,fig:ccfg_visual_2} show that there is no significant difference between the methods.
Importantly, decoding the bit-streams produced by CCFG does involve accessing the original input $\rvy$, and so our loss in CLIP scores are expected due to the rate-perception-distortion tradeoff~\citep{pmlr-v97-blau19a} (here, we achieve $\frac{1000\cdot\log_{2}(64)}{768^2}\approx 0.01$ BPP).
Note that using CCFG in DDCM is fundamentally different than using CFG (\Cref{sec:method}), since the latter requires access to $\rvy$.
\clearpage
\subsection{Compressed Text-Based Image Editing}\label{app:editing}

Image editing has seen significant progress in recent years, particularly in solutions that rely on pre-trained DDMs.
Generally speaking, most methods rely on inversion~\citep{huberman2024edit,manor2024zeroshot,wu2023latent,wallace2023edict,hertz2023delta}, noising and denoising the original input~\citep{meng2022sdedit}, or manipulating the DDM's attention layers~\citep{hertz2023prompttoprompt,tumanyan2023plug}.
We demonstrate a simple approach for editing images using DDCM with a text-conditional DDM. 

Specifically, we compress an image while feeding the DDM with a source prompt $c_{\text{src}}$ describing the image.
To edit the image, we start by decoding it up to timestep $T_\text{edit}$ while feeding the DDM with the original prompt $c_{\text{src}}$.
We then continue with the decoding from timestep $T_\text{edit}$, while feeding the DDM with the target prompt $c_{\text{dst}}$.
Intuitively, this approach should preserve the low frequency contents in the original image (e.g., objects), while guiding the image towards the described edit.

We compare our approach to DDPM inversion~\citep{huberman2024edit} and DDIM inversion~\citep{song2021denoising}. 
For DDPM inversion we set $T=100, T_\text{skip}=36$, as recommended by the authors. For DDIM inversion we set $T=100$, and follow \citet{huberman2024edit} to additionally apply DDIM inversion mid-way, using $T_\text{skip}=40$.
For our approach, we set $T=1000$ and $T_\text{edit}=600$. Note that this value of $T_\text{edit}$ is equivalent to $T_\text{skip}=40$ if $T=100$ sampling steps were used. 
For all methods we use the pre-trained Stable Diffusion 1.4 checkpoint from \href{https://huggingface.co/CompVis/stable-diffusion-v1-4}{Hugging Face}. Images are taken from the modified-ImageNetR-TI2I dataset~\citep{huberman2024edit, tumanyan2023plug}.

\begin{figure}
    \centering
    \includegraphics[width=1\textwidth]{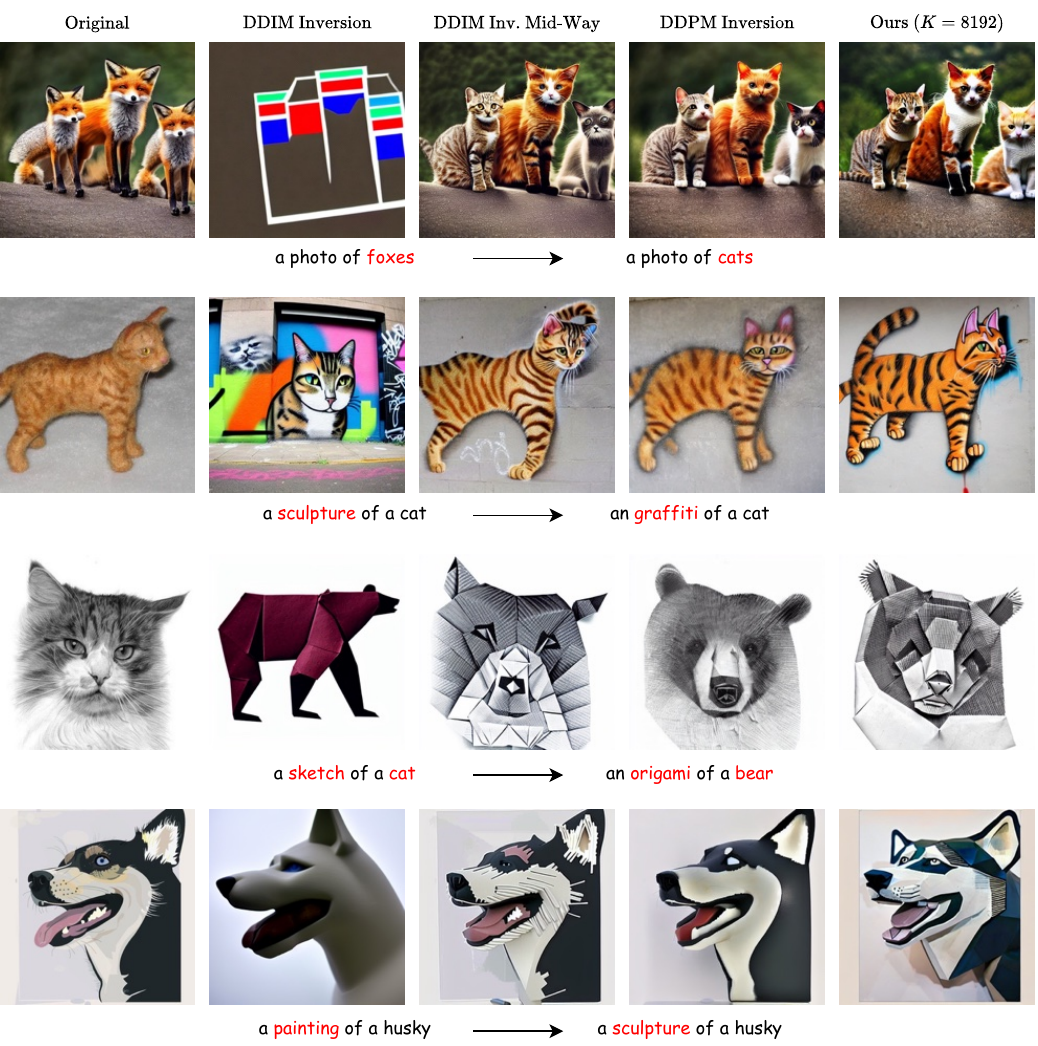}
    \caption{\textbf{Qualitative comparison of image editing methods.} Our approach preserves less image structure compared to DDPM inversion, while offering more semantic object preservation than DDIM inversion. This can be quite useful in scenarios where the editing prompt requires major structural changes, such as transforming a sketch of a cat into an origami of a bear.}
    \label{fig:editing}
\end{figure}

Preliminary qualitative results are shown \Cref{fig:editing}.
Our approach is less structure preserving than DDPM inversion, while offering more semantic object preservation than DDIM inversion. 
We encourage future works to investigate DDCM-based image editing methods.

\end{document}